**Title**
# Learning constructive models of emergent systems

**Authors**
Bipul Pandey[1], Caden Proctor[2], Vinay Ramanathan[3], Nico Roth[3], and Arjun S. Raman[1,3,4,5]

**Affiliations**
[1]Department of Pathology, University of Chicago, Chicago, IL, 60637
[2]Genomics, Genetics, and Systems Biology, University of Chicago, Chicago, IL, 60637
[3]Duchossois Family Institute, University of Chicago, Chicago, IL, 60637
[4]Pritzker School of Molecular Engineering, University of Chicago, Chicago, IL, 60637
[5]Center for the Physics of Evolving Systems, University of Chicago, Chicago, IL, 60637

## Abstract

Emergent systems – systems with multiscale interactions – arise through iteration than forward design, leaving principles for building them under-explored. Current artificial intelligence architectures, while useful for generation, provide no logic for construction. Inspired by statistical phylogenetics, we show that inferring scales of system entropy then constraining information from low to high entropic scales yields stepwise constructive models. We term this architecture the Layer-Restricted Boltzmann Machine (LRBM). Applied identically to digit images, natural language, and enzyme sequences, LRBMs follow a common hierarchical logic: lower layers encode global structure, higher layers fine-grained features. As a stringent test, we assayed 160 synthetic chorismate mutase (CM) constructive trajectories alongside 1,130 natural homologs *in vivo*. Designed CMs functioned up to 53% divergent from nearest natural homolog. Fold emerged at intermediate layers while function required all layers, illustrating that fold was necessary but insufficient for function. Our results demonstrate a shared, learnable logic for constructing emergent systems.

**One-sentence:** A constructive logic for producing synthetic emergent systems — from language to proteins — can be inferred as a hierarchy of entropic scales, suggesting a general architecture to the organization of emergent systems.

## Main

A foundational ambition of science is to elucidate principles underlying systems whose detailed mechanisms are too complex to follow directly. Statistical physics offers a paradigmatic example: rather than tracking every molecular trajectory, it enables an understanding of macroscopic organization by identifying constraints on degrees of freedom defining the space of possible system states. This same challenge and opportunity arise across biology, ecology, and the social and technical sciences where collective structure and function emerge from complex, multi-scale interactions amongst many heterogeneous components that comprise systems of interest[1–8]. We now possess extensive data regarding composition of such 'emergent' systems across these domains — sequenced biological systems, large text and image corpora, catalogued populations — and with it the means to create statistical representations enabling system generation[9–14]. This brings the search for general organizing principles within reach at least in the form of description: the creation of models that learn the space of possible states that systems can occupy.

Modern machine learning (ML) and artificial intelligence (AI) frameworks are a manifestation of this goal. Generative statistical models — Potts models, Boltzmann machines, variational autoencoders, diffusion models, transformers, and newer statistical architectures — learn the capacity to sample from a distribution of viable configurations that define an ensemble of systems[15–23]. In doing so, they can represent emergent systems in its strongest form to date: presented with a corpus of proteins, sentences, images, or other types of data defining related emergent systems, generative models learn a statistical representation from which new, plausible instances can be drawn. However, despite their obvious power, generative models lack the capacity to provide a recipe for constructing emergent systems. Although some generative models (diffusion, autoregressive models) proceed through an explicit order, that order is imposed by the method rather than inferred from the data[19,20,24]. Elucidating a statistical representation of how the ensemble of systems was built solely from observed data would represent a significant advance both practically and fundamentally. Practically, ‘constructive’ rather than generative modeling would (i) distinguish system components that are foundational from those that are incidental, (ii) provide a rationale for system architecture and perturbation, and (iii) offer a controlled route to constructing synthetic systems outside what has already been observed. Fundamentally, a model of the constructive process would be a substantially more general representation than merely a statistical description of its outputs. An accurate constructive model would specify constraints that specific instances of a family of systems must satisfy and therefore delimit the space of systems that could exist.

One reason that building constructive models of emergent systems is challenging is lack of access. A generative model can be learned because the systems it ultimately represents are reflective of the training data. In contrast, the process that builds systems is generally inaccessible to us as observers because this process primarily proceeds through iterative ‘tinkering’. This phrase, originally coined by Francois Jacob, references cycles of variation and selection defining the evolutionary process where what is ultimately observed are natural systems in the form of extant diversity[25]. The principle of engineering through iteration is not necessarily unique to biological evolution: physical artifacts, language, and other systems that evolve over time to come to be are shaped by reuse, variation, and recombination. In general, the constructive procedure underlying this iterative design is mostly unavailable for modeling but is nonetheless critical to understanding the current form of these systems. This raises a basic question: is any type of constructive logic possible to infer from access to only the ensemble of resulting systems?

The study of phylogenetics in Biology offers inspiration for addressing this problem. Living systems have come to be through iterative variation and selection that progressively refines and elaborates their organization over evolutionary time. A rich history of the study of phylogenetics from a statistical perspective has demonstrated that the relationships amongst extant organisms can be reconstructed from extant diversity alone without observing the process that produced this diversity[26–30]. Recent work has shown that this reconstruction can be performed by statistical inference directly: hierarchical relationships among biological systems and their constituent components are recoverable from the statistical structure of extant variation without being supplied phylogenetic information *a priori* or assumptions regarding patterns of genetic interactions and rates of molecular evolution[31,32]. These efforts provide a direct hypothesis that we evaluate in this work. Hierarchical scales of organization are present within the statistical structure of the extant diversity of emergent systems, and these scales enable the creation of constructive statistical models: a hierarchically ordered sequence of inferred constraints, imposed from coarse to fine, that serve as statistical rules of assemblage for emergent systems.

Here, we show that constructive models of emergent systems can be built from observed system diversity alone using only two operations with no additional architecture, statistical assumptions, notion of system dynamics, or domain knowledge applied. The first operation is to infer discrete scales of entropy from an ensemble of extant diversity where the scales collectively define a progression of constraints from those shared across the whole ensemble (e.g. low-entropy constraints) to those that distinguish unique aspects of individual system structure (e.g. high-entropy constraints). The second is to impose a single direction of information flow across these scales to train a statistical model: each scale is conditioned only by the scale immediately coarser than itself. These operations define a hierarchically layered model architecture we term the ‘Layer-Restricted Boltzmann Machine’ (LRBM).

Applied identically across several different classes of emergent systems — images of digits, natural language, and enzyme sequences — the LRBM leverages coarse-to-fine entropic scales and derives a hierarchical logic by which synthetic systems are constructed *de novo* from their constituent parts. In each case the LRBM is directly interpretable, computationally sparse, and able to synthesize new systems accurately through building them in a stepwise manner. For the case of enzyme construction, we synthesized 800 synthetic chorismate mutase enzymes spanning 160 constructive trajectories — each a random starting sequence evaluated through each LRBM layer — and assayed them alongside 1,130 natural homologs for *in vivo* complementation within a bacterial selection system. The final constructed sequences recovered both the native fold by prediction and intrinsic bimodal distribution of activity found across natural enzymes by experiment. Notably, we found that predicted fold and function were fixed at different stages of construction. Lower LRBM layers specified fold, including active site regions and the hydrophobic core, while functional activity emerged only upon satisfying the highest layers. Thus, satisfying fold was necessary but insufficient for function in this enzyme family, directly illustrating the worth of constructive models within a case study of statistical protein design.

Together, these results reveal a common logic of construction across disparate classes of emergent systems — images, language, and biological material — that share no obvious underlying mechanism of assemblage. In each, the same two operations, (i) inferring hierarchical entropic constraints and (ii) enforcing unidirectional information flow through this hierarchy, yield constructive models of emergent systems without domain-specific modification or information regarding the true underlying constructive process. Our work suggests that hierarchically related entropic scales may be a general organizing principle for learning the architectural logic of emergent systems *de novo*.

## Results

### *A hierarchical framework for constructive inference*

To build intuition about statistically inferring constructive processes of systems, we began by considering a scenario wherein an unknown generative process produces an ensemble of related systems. The resulting ensemble can be defined as a table of systems (rows) by features (columns) (**Fig. 1A**). We then defined a modeling process that considers the resulting systems as input and captures potentially multiscale interactions amongst system components. Our process is the following. First, we decompose the information content of the ensemble into discrete scales of entropic organization. This decomposition defines a hierarchy in which lower scales of entropy capture global constraints while higher entropic scales encode progressively variant system features. Second, we sequentially integrate these scales statistically: starting from the lowest entropic layer, we allow the flow of information only between adjacent layers (e.g. Scale 1 to Scale 2, Scale 2 to Scale 3, and so on) and explicitly prohibit backward connections (e.g. Scale 3 to Scale 2) (**Methods**). Given a set of inferred scales $\{S_1, \dots, S_i, \dots, S_K\}$, the statistical probability distribution of extant states within this framework is

$$P_{extant} = P(S_1) \prod_{i=1}^{K-1} P\,(S_{i+1}|S_i) \qquad (1)$$

Equation 1 plus constraining information flow in a hierarchical unidirectional manner defined the basis for training our constructive model. The resulting model is hierarchical, containing the same number of layers guiding system construction as scales of entropy. We note that our process inverts the standard paradigm of model building: rather than inferring system organization from generated outputs, we use entropic organization inferred from the data itself to define and constrain the model. In this view, inferred statistical structure is actively used as a feature for model creation. In all results shown subsequently, we used **Eq. 1** to train a Boltzmann Machine with this inferred structure and therefore termed the resulting architecture the Layer-Restricted Boltzmann Machine (LRBM).

To determine whether this framework could approximate an underlying generative process consisting of complex constructive logic, we devised an evolutionary simulation where we defined a generative process to create an ensemble of diverse systems and trained an LRBM based on the resulting ensemble only (**Methods**). This simulation was defined by two parts: (i) a fitness function to determine how the population of systems would be iteratively shaped through rounds of selection and (ii) the process of selection and variation. With regards to the fitness function, in our simulation each system was defined by 16 positions, and each position could be one of four possible states—A, C, T, or G. We defined the 'fitness' of a system across four discrete layers where positions defined by the first two layers carried the greatest fitness importance, positions defined by the third layer were relatively less important, and the fourth layer was neutral (**Fig. 1B**, left). In this way, resulting systems were multiscale with respect to the relative importance of their constituent components, similar to an evolutionary phylogeny. With regard to the process of selection and variation, the population size of the simulation was held constant at 5,000 and the composition of the population changed during each epoch. Within each epoch, a set of steps was used to define selection and variation (**Fig. 1B**, right). First, all systems with fitness equal to zero were purged from the population. Second, systems were then mutated at random; this population filled in the empty niche. Third, a random subpopulation was chosen where pairwise competition between systems was performed such that the higher fitness system would be selected in a probabilistic manner (not deterministically). The losing population was purged; the winning population was maintained and allowed to 'propagate' with random mutation. Notably this pairwise competition step may enable competition between two high fitness or low fitness systems against each other. As such, the probabilistic nature of selection allowed the possibility of a lower fitness system to stochastically be positively selected over a higher fitness system. These two considerations enabled maintaining a distribution of fitness values rather than simply converging in a pseudo-deterministic limit to uniformly high fitness systems. Our simulation closely followed previously described Moran-type evolutionary algorithms[33,34,55] (**Methods**).

This simulation was carried out for 10,000 epochs, resulting in the ensemble of system diversity that we used to train an LRBM. Notably, the generative logic of our simulation directly mirrored the mathematical structure of the proposed framework in **Fig. 1A** where variation is introduced through a sequence of conditional dependencies across discrete entropic scales. As such, this simulation provided a minimal and controlled setting in which the assumptions of the model were exactly satisfied and thereby constituted a necessary baseline: if the LRBM model could not recover structure in this setting, it could not be expected to recover structure in natural systems.

We first computed the scales of entropy in the ensemble of sequences produced by the simulation. We found that there were four discrete scales of entropy revealed from statistical inference performed on the ensemble of resulting systems. Additionally, these scales and their relative entropy precisely corresponded to the four hierarchical layers of constraints imposed on the generative process described above (**Fig. 1C**, left). This result demonstrated that computing entropic scales of organization from the output of the generative process revealed the underlying organization of generative variation without access to the generative process itself. We next imposed this four-cluster structure and trained an LRBM (**Methods**). By construction, the LRBM model parameter matrix, described by couplings between positions, exhibited a block-structured organization reflecting unidirectional dependencies across entropic scales (as described by **Eq. 1**) (**Fig. 1C**, right). The coupling strength was strongest amongst positions contained within scale 1—the lowest entropy scale—and progressively declined such that interactions amongst positions contained within scale 4 exhibiting the least coupling. This structured decay in coupling strength recapitulates the hierarchy of constraints encoded in the generative process: low-entropy scales exert dominant, global influence over system organization and higher-entropy scales contribute progressively weaker, localized variation nested amongst low-entropy scales.

We next compared the performance of our trained LRBM to that of a standard profile-based model[58] that captures only single-site statistics (**Methods**). Analysis of 5000 LRBM-generated systems illustrated three notable results. First, the LRBM produced novel systems as judged by non-zero hamming distance from the most closely related system in the training data (**Fig. 1D**, top left). Second, the LRBM reproduced statistical dependencies across individual, pairwise, and triplet frequency distributions as well as higher-order correlations faithfully (**Fig. 1D**, top middle). Third, the LRBM produced a distribution of systems with fitness effects that recapitulated the fitness distribution of the training data (**Fig. 1D**, top right). While the profile-based model also produced novel systems and reproduced single-site frequency distributions across these systems, this model failed to capture higher-order dependencies between positions (**Fig. 1D**, bottom left and middle). As a result, profile-based generated systems were selectively unable to recapitulate the high fitness regime of the training data (**Fig. 1D**, bottom right). Capturing higher-order statistical structure was therefore required to access system compositions associated with high-fitness, and this structure was faithfully encoded by the LRBM but not by models limited to independent-site statistics.

Taken together, these results demonstrated that our statistical framework and the resulting LRBM could recover the hierarchical structure of an underlying generative process directly from its outputs within our simulation framework. Having established this correspondence in a setting where the generative rules are known and follow a specific hierarchical architecture, we next applied this framework to systems where we did not have access to the underlying generative process.

*LRBM-based construction in images and written language*

We applied the LRBM framework to the Modified National Institute of Standards and Technology (MNIST) dataset[67] of handwritten digits. This dataset provides a well-defined ensemble: each sample is a 28 x 28 image of a digit generated through human motor variability and a social acceptance of number recognition (**Fig. 2A**, top). As a result, the dataset contains rich, structured variation — global constraints such as digit identity and topology are coupled to local variation in stroke thickness, curvature, and orientation. Importantly, (i) we considered a total of 500 randomly chosen examples (50 for each digit going from 0 through 9) out of 70,000 total examples (ii) the underlying generative rules of these digits are not explicitly known, and (iii) pixels collectively define the digit thereby making digit identity a collective, emergent property of interaction amongst components. Although MNIST dataset includes digit identity labels, we did not use these labels for training or generative purposes.

We represented our MNIST dataset corpus as a matrix of samples by features, where each image was binarized such that individual pixels were treated as ON/OFF variables. Thus, the total combinatorial complexity of this dataset is $2^{784}$ or $10^{236}$. In this representation, each row corresponded to an MNIST digit and each column a pixel (**Fig. 2A**, bottom). We then decomposed this ensemble into entropic scales following Eq. 1 (**Methods**). We found that the spatial representation of these scales was structured across the physical space of pixel placement (**Fig. 2B**, top). The lowest entropy scale captured the global constraint separating white space from pixels that could be occupied as part of a digit; progressively higher entropy scales refined this description, focusing the allowable pixel-space to be higher in resolution. This decomposition therefore resolved the images into hierarchical scales of constraint directly from data without reference to spatial priors or labels. Using this decomposition, we trained an MNIST-LRBM (**Methods**). The resulting model accurately recapitulated the statistical structure of the dataset across multiple orders, including individual pixel frequencies as well as pairwise and higher-order correlations (**Fig. 2B**, bottom).

To test whether the MNIST-LRBM defined a constructive process, we next examined whether digits could be reconstructed through sequential integration of entropic layers. Starting from a corrupted input in which 50% of pixels were randomized, we projected the noised image through the LRBM (**Methods**). Early LRBM layers, encoding low-entropy constraints, recovered coarse global structure and reinstated the presence and approximate topology of the digit within the image. Subsequent layers incrementally refined this representation, resolving stroke placement, curvature, and local pixel configurations (**Fig. 2C**). This stepwise reconstruction demonstrated that each entropic layer contributed distinct ordered information such that accurate recovery of a coherent digit emerged only after integrating the full hierarchy. Notably, the final reconstruction was not an exact replica of the original input, but rather a configuration that satisfied the hierarchical constraints inferred from the data. In this sense, progressing a noised image through the MNIST-LRBM did not simply recover an instance of the training data.

We next sampled directly from the MNIST-LRBM, generating synthetic images of handwritten digits from purely random noise input. Generated digits exhibited a broad distribution of Hamming distances from the training set, indicating that the MNIST-LRBM did not memorize inputs (**Fig. 2D**, left). Rather, samples populated configurations extending beyond the typical variation observed in the training data. Despite this increasing distance, generated images remained structured and interpretable. At moderate distances above the mean (e.g. Hamming distances of d = 57 and d = 73), samples corresponded to coherent digits such as a ‘2’ and ‘3’ respectively (**Fig. 2D**, right). This demonstrated that global structure was preserved while local features nested within global structure diverged. At larger distances (e.g. d = 92), samples adopted hybrid morphologies consistent with multiple different digits (e.g. a ‘5’ or ‘6’), reflecting a satisfaction of broad structural requirements to be a digit. Taken together, these results demonstrated that hierarchical constraint structure could be inferred from MNIST and used constructively to generate valid, non-memorized digits.

We next sought to determine whether the LRBM framework could generalize to language — a system where structure is not spatially organized but emerges across sequences of discrete symbols. We collected 1,710 sentences from Simple Wikipedia animal articles (**Fig. 3A**, left) (**Methods**). After removing all non-alphanumeric characters and converting numerals to words, 2,855 unique words were identified. We recursively split any word longer than three characters if a contiguous sub-word existed elsewhere in the dataset, yielding the set of tokens representing the data (**Fig. 3A**, middle and right). We then added each new token back to the list of words and repeated the process until no further splits were possible. A small set of common words (e.g., ‘they/them’ converted to ‘the’ + ‘y/m’) were protected from splitting. This recursive tokenization reduced vocabulary by 24.5% (2,855 to 2,156 unique words) while making grammatically meaningful relationships like pluralization ("cows" became "cow" plus "s"), noun-adjective conversion ("American" became "America" plus "n") explicit. We then filtered these sentences to a maximum of 17 tokens retaining 1,589 sentences, front-aligned them with blank end-paddings, and computed per-position mean and deviation of entropy via 200-bootstrap sampling. The resulting complexity of the dataset was $2{,}156^{17}$ or roughly $10^{56}$.

There existed four entropic scales in our corpus of sentences (**Fig. 3B**, top) (**Methods**). The first scale comprised sentence positions 11-17 and was reflective of the end-padding alignment we had performed, leading to blank spaces at these positions in the corpus. This scale thereby represented the distribution of sentence lengths in the input data (**Fig. 3B**, right and bottom). The remaining scales were nested with positions 4-7 forming the highest entropy scale reflecting a distribution of possible verbs (**Fig. 3B**, bottom). The degeneracy of possible verbs suggested a multiplicity of actions given a fixed sentence subject and object specified in the earlier scales. We next created a Simple Animal LRBM (SA-LRBM) using a color-compressed framework (**Methods**) and generated 2000 sentences *de novo*. Generated sentences showed string preservation of single-site and pairwise frequencies (Pearson r = 0.97 and 0.95, respectively) as well as preservation of pairwise correlation (Pearson r = 0.61) (**Fig. 3C**).

To evaluate whether generated sentences were consistent with natural language, we computed the LRBM energy for all natural and generated sentences and examined their distributions across the four scales of entropy (**Fig. 3D**) (**Methods**). When sampling only from the lowest-entropy scale, generated sentences were mostly aligned by sentence length corresponding to the structural scaffold of the sentence. Inclusion of higher-entropy scales introduced grammatical constraints and semantic coherence, with sentences becoming increasingly structured. By satisfying all four entropic scales, the SA-LRBM produced sentences with recognizable sentence structure.

We next evaluated whether the SA-LRBM was learning constraints on word position consistent with grammar or merely the presence or absence of specific words. To address this, we changed the order of word positions within sentences while preserving the choice of individual words and then computed SA-LRBM associated energies of each sentence. Our rationale was that if the SA-LRBM learned grammatical constraints, the energy of sentences would become less favored as the order of words was scrambled while the individual words remained the same. We computed SA-LRBM energies for all sentences where we changed the order of the words in a sentence but kept the word choice the same. We did sentence scrambling for as few as two positions and up to 11 positions. We found that even swapping the position of two words in a sentence resulted in substantially higher energy values. Energy values continued to increase as the number of words that were swapped increased to 8 and reached an asymptotic energy level from 9 to 11 swaps (**Fig. S1A**, left). As an example, a natively generated sentence was '*This means they are awake at night and asleep during the day*' and had an associated SA-LRBM energy value of -142.82. Swapping two words resulted in the sentence '*This means are they awake at night and asleep during the day*' with an associated SA-LRBM energy value of -129.21. Swapping 10 words resulted in the sentence '*This means are asleep during the night day they at and awake a*' with an associated SA-LRBM energy value of -82.97 (**Fig. S1A**, right). These findings illustrated that the SA-LRBM learned contextual relationships between words, suggesting that the SA-LRBM energy reflected the semantic meaning of sentences.

Our results so far had demonstrated that the LRBM architecture suggested an ordered logic of construction: each layer successively constrains which system configurations are possible and therefore the hierarchical structure reorganized the space of possible sentences into a layered coordinate system. If true, we reasoned that melting constructed sentences at different layer depths should reveal how each layer partitions the landscape. Specifically, coarse layers would define broad regions of possible sentences while finer layers would progressively narrow the search space to specific instances. To address this idea, we (i) started with a natural sentence, (ii) melted the sentence in a layer-specific manner (Layer 4, Layer 3 and 4, Layer 2 through 4), and (iii) rebuilt sentences by cooling and satisfying the remaining layers. Using the starting sentence of '*This means they are awake at night and asleep during the day*', we first melted Layer 4 and rebuilt sentences. Reconstructed sentences maintained consistent semantic structure: all sentences began with '*This means'* and ended with '*asleep during the day*', indicating that melting only the finest layer kept the sentences within a constrained neighborhood of the space of language (**Fig. S1B**, top). After melting and reconstructing sentences according to Layers 3 and 4, sentences showed greater variability in semantic content and grammar, spanning a broader region of language space (**Fig. S1B**, middle). When we melted and rebuilt sentences according to Layers 2 through 4, reconstructed sentences jumped to entirely different semantic neighborhoods such as '*Pigs are often called the mammals and asleep during the day*' (**Fig. S1B**, bottom). These results revealed that the hierarchical architecture of the LRBM organized the space of system configurations as nested, concentric regions: melting only fine layers kept exploration local within a neighborhood, while melting progressively coarser layers allowed access to entirely different, yet acceptable, regions of language space. Thus, the LRBM layers structured the space such that different melting depths could enable exploration across semantic content from one solution to another.

Given the grammatical simplicity of *'Simple Wikipedia'* corpus, we next asked whether we could create a constructive model for another corpus of text data. We tokenized all 748 sentences in the novel *'Alice in Wonderland'* by Lewis Carroll and restricted the corpus to 716 sentences of length 17 comprising of 799 unique tokens, computed scales of entropy, and created a trained LRBM (AW-LRBM). The total combinatorial complexity of this dataset was $799^{17}$ or roughly $10^{49}$. We found four scales of entropy corresponding to a constructive logic that the resulting trained LRBM followed (**Fig. S2A**, **B**). Generated sentences reflected the structure and content of sentences within the training data (e.g. '*And she hadn't seen the knave about the size of the white rabbit's house*') (**Fig. S2C**). Moreover, we found that the resulting LRBM again learned grammatical structure rather than word choice (**Fig. S3A**, left). As an example, a low energy AW-LRBM constructed sentence was '*Well you see it had to go and visit the duchess*'. Scrambling 8 words resulted in a substantially higher energy score and the sentence '*To you see it had duchess go and well the visit*' (**Fig. S3 A**, right).

We next repeated the experiment where we melted and cooled sentences according to the AW-LRBM architecture. Like the SA-LRBM, we found that melting then cooling sentences according to the AW-LRBM layers progressively explored more of the language space encoded by the *'Alice in Wonderland'* corpus. As an example, given the constructed sentence '*Well you see it had to go and visit the duchess*', melting the Layer 4 AW-LRBM information then reconstructing sentences maintained '*Well you…go and visit the duchess'* and locally explored the language space, thereby resulting in sentences like '*Well you <u>knew you wouldn't go</u> and visit the duchess*' (**Fig. S3B**, top). Melting the sentence according to AW-LRBM Layers 3 and 4 and then reconstructing the sentence resulted in maintenance of '*Well…visit the duchess*' and more globally explored the language space, resulting in sentences like '*Well Alice hadn't seen the queen visit the duchess*' (**Fig. S3B**, bottom).

Together, the MNIST and language results illustrated that the LRBM framework generalizes across different types of data — spatial images defined by collections of local pixel interactions as well as qualitatively distinct corpora of language data. Despite these seemingly profound differences in class of system, the same operations of (i) inferring entropic scales of organization and (ii) imposing unidirectional information flow resulted in constructive models. Moreover, these results demonstrated that the LRBM can construct hierarchical models reflecting local, corpus-specific data without requiring internet-scale pretraining. We next sought to determine whether our constructive framework could extend to biological systems where the generative process is governed by evolution.

*<u>Constructive design of synthetic chorismate mutases</u>*

Extant protein sequences reflect the outcome of iterative cycles of variation and selection associated with survival. As a result, ensembles of sequences belonging to a given protein family are sculpted in a sequentially iterative manner by phylogenetic history. Numerous studies have shown that evolutionary sequence diversity encodes statistical dependencies between amino acids that are informative of protein structure, function, and adaptability to new function[35–40]. As a different statistical strategy, other recent studies have embraced deep learning approaches to generate functional proteins directly from sequence data[41–43]. Current protein design approaches do not learn a constructive model of protein design — a stepwise order of particular amino acids within particular residues to ultimately build a functional protein. We evaluated whether we could use our framework to achieve this goal.

As a model system, we focused on the enzyme chorismate mutase (CM). This enzyme catalyzes a key step in aromatic amino acid biosynthesis required for cellular growth[44,45]. We selected enzymes as a stringent test case for protein design because unlike structural or scaffolding proteins, the function of enzymes is thought to depend on the precise coordination of multiple constraints — global fold, active site geometry, and catalytic chemistry — making them a non-trivial target for protein design. We selected the CM family because there has been extensive prior characterization of these enzymes, including the creation of generative models from evolutionary statistics[46,47].

We first used a previously created multiple sequence alignment (MSA) of the chorismate mutase (CM) family comprising 1130 natural sequences (**Fig. 4A**, left; **Table S1A**)[46]. From this MSA, we computed scales of entropy present in the data (**Methods**). In total, there were five scales. The first two scales specified residues that were previously found to be involved in catalysis — Arg11, Arg28, Lys39, Arg51, Glu52, Gln88 — as well as specific amino acids known to interact with active site residues in a functionally relevant manner — Ile14, Val35, Leu55, Ile81, Ser84 (**Fig. 4A**, middle)[48]. The positions specified by the first two scales comprised amongst the most conserved positions within the family. These included many positions that constituted the hydrophobic core of the protein as well as the positions directly involved in the catalysis of the substrate (**Fig. 4A**, middle). Scales 3 through 5 progressively specified residues away from the catalytic core (**Fig. 4A**, middle; **Fig. S4A**). From these scales, we trained an LRBM — CM-LRBM — resulting in a five-layer constructive model where a random seed sequence could be progressively sculpted into a sequence satisfying all five CM-LRBM layers (**Fig. 4A**, right).

The CM-LRBM accurately reproduced the statistical organization of natural sequences across single-site frequencies, pairwise frequencies, pairwise correlations, and higher-order triplet correlations, matching the training data with high fidelity (**Fig. 4B**, left). We next generated 2,000 synthetic CM sequences by sampling from the fully trained CM-LRBM where all five layers were satisfied and selected 160 of them to evaluate their function using an *in vivo* complementation assay in *Escherichia coli*. These sequences matched the frequency distributions of amino acids within the training data (**Fig. S4 B**). Within this assay, bacterial growth was directly coupled to catalytic activity of the enzyme (**Fig. S4 C, D**) (**Methods**) (**Table S2A**). We also included a 'null' CM that was a sequence 55% of the length of a full CM and therefore served as a negative control in our experiment. We synthesized and assayed the 160 sequences in addition to the null and all 1,130 natural CMs in the MSA. We found that the 1,130 natural sequences demonstrated a bimodal distribution of enzyme function with many sequences that were non-functional and others that were nearly as functional as the CM for *E. coli*. We found that the 160 synthetic sequences accurately captured this bimodal distribution of enzyme function (**Fig. 4B**, right; **Fig. S5 A, B**) (**Table S2B, S2C**). Moreover, analysis of the synthetic sequences illustrated that functional sequences were deviant from both the nearest natural homolog — up to 53% sequence divergent as well as from the *E. coli* CM – up to 75% sequence divergent (**Fig. 4C**) (**Table S2C**). These results demonstrated that the CM-LRBM created an ensemble of synthetic proteins that were distinct from the training data and recapitulated the distribution of enzymatic function across the set of natural CMs.

Given that the CM-LRBM could generate functional synthetic chorismate mutase enzymes, we next interrogated the layers of the CM-LRBM with respect to the biological information they encoded. We synthesized all constructive trajectories layer-by-layer for each of the 160 sequences we had created (a total of 640 sequences) and evaluated all sequences for their catalytic activity using our *in vivo* complementation assay (**Table S2D-G**). Each trajectory started with a random sequence and was progressed through each of the five CM-LRBM layers. We set out to test 160 complete trajectories, totaling to 800 synthetically generated sequences. After sequencing, we filtered out sequences with less than 10 input reads to avoid noise in our enrichment calculations. This left us with 759 quality-controlled datapoints, representing 120 complete trajectories. We computed the CM-LRBM energy of each sequence from the fields provided by the layers of the trained model (**Methods**). The null CM was high energy with minimal activity. Sequences satisfying only Layer 1 clustered near the null control at high LRBM energy with minimal activity. Successive addition of layers up through Layer 4 progressively decreased LRBM energy but still achieved relatively minimal activity (**Fig. 5A**; **Table S2D-G**). The distribution of CM-LRBM energies for each layer of the 160 experimentally tested proteins matched that of the 2,000 constructed sequences used to evaluate the fidelity of model training, demonstrating that our experimentally tested proteins were reflective of the energy landscape represented by the CM-LRBM (**Fig. S5C**). Thus, despite catalytic residues and the hydrophobic core being specified within Layers 1 and 2, satisfying intermediate CM-LRBM layers was sufficient to generate active enzymes.

We next evaluated the fold quality of all 800 constructed sequences as well as the set of 1,130 natural CM sequences used for LRBM training using two independent structure-prediction models: AlphaFold3 (AF3) and Rosetta[11,49]. For each of the sequences constructed in the trajectories as well as the natural sequences, we computed per-residue and structural confidence scores resulting from AF3 (pLDDT, ipTM, pTM) and structural stability scores resulting from Rosetta (Rosetta Binding Energies, Steric Clash, Total Energies) across all layer levels (**Fig. 5B**, **Fig. S6A-C**; **Table S1A-F**). Layer 1 sequences showed poor confidence structure prediction – low pLDDT, and high Rosetta Binding Energies – indicating unfolded or severely misfolded structures. Successive satisfaction of Layers 2 and 3 progressively improved fold quality, with confidence scores improving and matching the confidence of natural CMs thereby indicating well-folded structures. Critically, confidence scores saturated by Layer 3: Layers 4 and 5 showed no further improvement in confidence scores across both AlphaFold3 and Rosetta predictions, indicating that the native-like fold had likely been fully established by Layer 3 and remained stable through the remaining layers. Same trends were observed in all other AF3 and Rosetta metrics (**Fig. S6**). Consistent with this finding, the mean Cα root mean square deviation across predicted structures for all synthetic proteins was uniformly high at Layer 1, began to stabilize within the hydrophobic core at Layer 2, and was uniformly low at Layer 3, where synthetic proteins resembled that of natural sequences (**Fig. 5B**; **Table S1G**) (**Methods**). This result revealed a biological decomposition encoded by the CM-LRBM: the catalytic machinery was assembled by Layer 2, the enzyme fold was assembled by Layer 3, and Layers 4 and 5 refined primarily surface-exposed sites necessary for achieving catalytic activity. The hierarchical, constructive architecture of the CM-LRBM thus revealed that fold is necessary but insufficient for function in CM.

The LRBM architecture provided the unique opportunity to visualize the stepwise assembly of a functional enzyme from both sequence and structural perspectives. We traced a single constructive trajectory from the seed sequence associated with the synthetic CM that was the most sequence divergent amongst our designs, exhibiting 47% sequence identity to the nearest homolog after satisfying all layers of the CM-LRBM. From the perspective of primary sequence, we found that the key residues of the enzyme related to catalysis — Arg11, Arg28, Lys39, Arg51, Glu52, and Gln88 — were specified at Layer 1 while Ile81 and Ser84 were specified at Layer 2. Leu55 was specified at Layer 2 but mutated to a valine, signifying a substitution to an amino acid with a non-polar but smaller sidechain. We noted however that these positions were not the only positions that were specified; numerous other positions were altered from the seed sequence to Layer 1 and from Layer 1 to Layer 2 (**Fig. 5C**, top). These residues were not obviously related to secondary structural elements and were primarily found within the hydrophobic core of the enzyme (**Fig. 5C**, bottom). As the trajectory progressed through Layers 3 through 5, the patterns of changes in the sequence and structure became increasingly distributed and difficult to interpret. Positions were modified at both buried and solvent-exposed sites with no clear organizing principle.

To understand whether this opacity between sequence, structure, and function was specific to a single trajectory or reflected a more systematic property of functional construction, we examined the amino acid composition at each position across the subset of 160 generated sequences that recapitulated the function of *E. coli* CM (n = 34) at each layer (**Fig. 5D**). At Layer 1, positions showed substantial but incomplete conservation. Key catalytic positions such as position 88 accepted variation amongst charged and polar amino acids rather than a single canonical residue. As successive layers were satisfied, the degeneracy of possible amino acids within specified positions increased. By Layer 5, numerous positions tolerated many kinds of amino acids. Notably, there was not an obvious amino acid signature at any layer that distinguished synthetic enzymes that functioned like *E. coli* CM and those that did not (**Fig. S7**). This widespread degeneracy illustrated that the information specifying catalytic function was not encoded in individual positions but in the collective, interdependent satisfaction of constraints within and across layers.

Collectively, our findings demonstrated three results. First, functional CMs can be constructed from the CM-LRBM. Second, the model reflected constructing the hydrophobic core and active sites, then fold, then function, illustrating that fold was necessary but insufficient for catalytic activity within this enzyme family. Third, the CM-LRBM energy metric captured functional specification in a way that sequence and structural decomposition could not: the sequence-based logic of enzyme function was opaque but notably, within the learned constraints of the CM-LRBM.

*Origins of novelty in LRBM constructive models and sparsity of compute*

Our results so far demonstrated that the LRBM generates novel systems across all four domains tested — our simulation, MNIST, language, and protein design — without memorizing the training ensemble. This motivated a more fundamental question: from where does novelty emerge? The architecture of the LRBM itself suggested a hypothesis. If entropic scales capture a hierarchy of constraints for building emergent systems, then coarse, lower layers should enforce tight constraint on system construction and permit little deviation from the training ensemble while fine layers that encode high-entropy variation should license greater freedom.

We tested this hypothesis by computing the proportional average Hamming distance of LRBM-generated systems to the training data at each partition level across all four domains (**Fig. 6A**). The hamming variability of components of the LRBM-generated systems associated with the most coarse, lowest-entropy scale and Layer 1 of the LRBM were low — similar to the corresponding training distribution. As the scales of entropy and LRBM-associated layers increased, the average Hamming distance also increased across all four domains. These results illustrated that the novelty in LRBM-generated systems primarily originated from higher LRBM layers nested amongst shared, more foundational constraints that were encoded within lower LRBM layers. The constructive architecture of the LRBM therefore provided a principled statistical mechanism for controlled exploration.

As the origins of novelty in the LRBM are layer-dependent, a natural next question was whether the LRBM hierarchies are consistent with known constructive processes by which the systems we interrogated were generated. Within the cases of MNIST and language, addressing this question was not feasible because of the lack of a 'ground-truth' classification with which we could compare constructed systems. However, in biological systems we have access to an independent record reflective of the underlying generative process: the phylogenetic hierarchy which documents an order in which evolutionary constraints were progressively established through a well-accepted classification scheme of Phylum to Species level designations. We therefore asked if the inferred entropic scales respect the multiscale phylogenetic structure of the CM family.

To test this idea, we created sequence based phylogenetic trees starting with the first entropic scale and progressively considered higher scales of entropy (**Fig. 6B**, **Figs. S8**, **S9, Table S3**) (**Methods**) mapping higher-level taxonomic classifications (kingdom through genus) directly onto the leaf coloration. We first noted that the inferred phylogenetic trees constructed from progressively increasing scales of entropy revealed a clear pattern. The deep interior nodes of the tree corresponding to well-structured, ancient differences remained topologically invariant across all scales of entropy. In contrast, the terminal branching structure, encoding more fine-scaled recent differences, progressively resolved as more scales of entropy were included to create the tree. The tree created only by the first scale of entropy established Kingdom-level separation with Bacteria, Archaea, and Eukarya occupying distinct regions of the tree. However phylogenetic mixing also persisted within these kingdoms. Notably, Fungi and Viridiplantae, both eukaryotes, were intermingled within the tree created by the first entropic scale. Considering entropic scales 1 through 3, this mixing resolved: Fungi and Viridiplantae separated into distinct clades. Similarly, considering entropic scales 1 through 5 also resolved within-kingdom mixing: *Bacillati* and *Pseudomonadati* became progressively better separated while they were entangled when considering coarser entropic scales. Overlaying our CM-LRBM generated sequences on these trees revealed that these sequences were not confined to a single lineage or pocket of sequence space, but rather populated the full evolutionary diversity of natural CMs in the MSA (**Fig. 6B**)(**Methods**). This demonstrated that the CM-LRBM captured the distributed evolutionary constraints for generating functional enzymes according to extant diversity rather than converging on a phylogenetically narrow set of solutions.

The correspondence between inferred entropic scales and evolutionary hierarchy illustrated that the constructive prior captured organizational structure amongst extant biological diversity. We observed that this structural alignment was associated with substantial computational efficiency of the LRBM models. The CM-LRBM comprised 1.24 million parameters occupying 38.9 MB; the SA-LRBM comprised 14.67 million parameters occupying 439.1 MB; the MNIST-LRBM comprised 855,346 parameters occupying 29.3 MB (**Table 1**). Across all systems, model size scaled in a way that was likely more associated with semantic or functional complexity, not with the combinatorial possibilities defining the underlying data spaces ($\sim 21^{96}$ for proteins, $\sim 2{,}156^{17}$ for language, $\sim 2^{784}$ for images). Training and inference were performed on local hardware – a 2023 MacBook Pro – without specialized accelerators or distributed resources. Combined with cross-domain generality, functional recovery, layer-indexed novelty, and phylogenetic alignment in proteins, these results demonstrate that the LRBM enabled constructive modeling with computational efficiency, interpretability, and the capacity for localized model creation without requiring internet-scale data or specialized infrastructure.

## Limitations

We note two limitations regarding our findings and framework. First, the LRBM was developed and tested exclusively on discretized data. Whether the framework generalizes to continuous variables, dynamical systems, or hybrid systems remains unknown and represents an important frontier in continued formal development. A second and more fundamental limitation is that our results serve as primarily a phenomenological discovery, demonstrating that the LRBM framework works across diverse domains but without a fundamental, statistical-mechanical or information theoretic understanding of why it works. We reason that the mechanism is rooted in constrained information flow across scales where the conditional distribution of each layer depends on layers that are entropically coarse compared to it. We note however that this architecture violates assumptions central to equilibrium statistical mechanics, where detailed balance and reversibility are presumed. Instead, the LRBM embodies fundamentally asymmetric, directional constraint flow, suggesting that emergent systems may be organized in ways that are more consistent with so-called 'non-Hermitian' systems where information flow is irreversible. Why this type of constraint architecture is sufficient to reconstruct functional systems, what theoretical principles govern it, and how it relates to deeper mathematical structures motivates a more formal treatment of constructive hierarchies.

## Discussion

We have demonstrated that emergent systems can be reconstructed as ordered hierarchies of entropy inferred from extant diversity alone, introducing the LRBM as a constructive architecture. The framework rests on two operations: (i) decomposing observed diversity into discrete scales of entropy via statistical inference, and (ii) imposing unidirectional information flow across these scales such that constructive modeling proceeds from low to high entropic scales. Our findings illustrate that just these two operations are sufficient to build functional systems across fundamentally unrelated domains — a simulated system with known ground truth, images of digits, written language, and enzymes — without domain-specific architecture, knowledge, or additional constraints. In the case of proteins, we demonstrated the value of constructive modeling by finding that the presence of functional (catalytic) machinery as well as fold is a prerequisite but insufficient on its own for building functional enzymes. As the LRBM architecture could be directly interrogated, we found that system novelty was layer-indexed and emerging from high-entropy scales, while low-entropy scales enforced constraint on system construction.

A fundamental principle in information theory and statistical learning is that when the structure of a model matches the organization of the data, the prior does substantial work that would otherwise require parameters and regularization. In standard generative models, the architecture is agnostic to data structure; variation and constraint must be discovered implicitly through high-dimensional optimization. The LRBM imposes structure explicitly: the entropy hierarchy constraints how information flows from coarse to fine scales, reducing the dimensionality of the solution space the model must explore. When such constraint is well-matched to the underlying system, the model should be parsimonious, capable of achieving its objective with fewer parameters and less computational overhead. Conversely, a drastic increase in computational parameters suggests a mismatch between model structure and data structure. This raises a hypothesis: the sparsity of the LRBM models across domains may be diagnostic of how well the hierarchical prior captures genuine organizational structure. This relationship between structural alignment and computational efficiency suggests that hierarchical priors in machine learning may yield benefits not merely in interpretability but also in efficiency and learnability. Systems organized according to hierarchical principles may be inherently more parsimonious than those approached as unstructured high-dimensional spaces.

Hierarchical modularity has long been proposed as a fundamental principle of organizing emergent systems. Previous work has put forth that emergent complexity arises through satisfying layered constraints where coarse organization is established first and successive refinement adds specificity without disrupting the foundation[1]. Experimentally evaluating these ideas has been historically challenging in most systems due to the immense number of possible component interactions that would need to be interrogated as we, the observers, do not usually have access to the generative process. Our work illustrates the existence of a statistical procedure that leverages variation amongst extant diversity to infer hierarchical modularity. Our results show that using this inferred hierarchical modularity effectively constructs emergent, complex systems and that intermediate layers of the constructive process are partially meaningful. In MNIST, recognizable digit structure emerges at middle layers; in language, grammatical scaffolding emerges progressively; in proteins, the fold is established prior to function. The generality of our framework across domains suggests that hierarchical modularity may underlie how emergent complexity is naturally built through the process of iterative refinement. If true, understanding the selective and physical criteria that enforce hierarchical modularity as a design principle of emergent complexity will be an important future endeavor.

## Figures

Pandey et al., Figure 1.

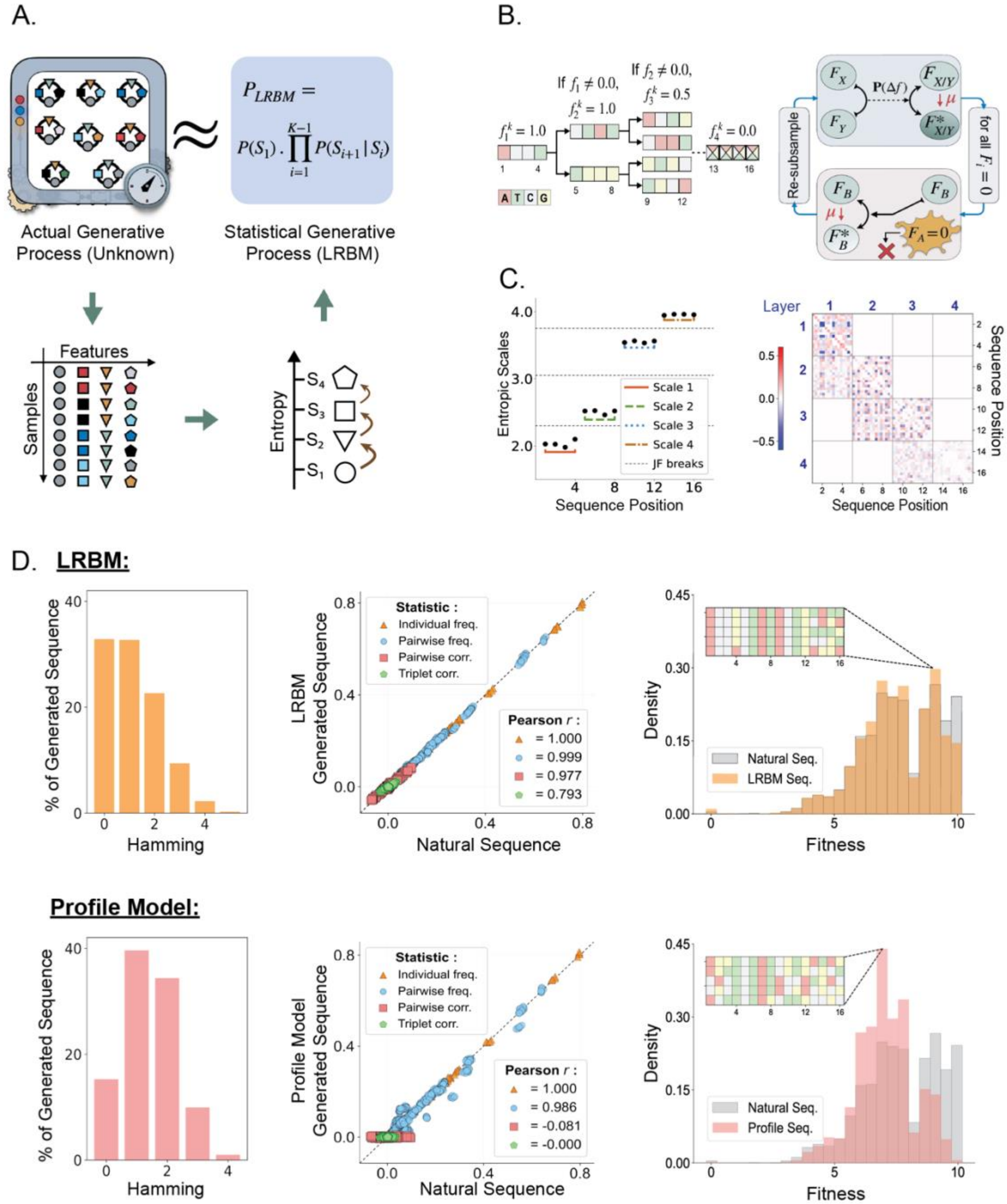

**Figure 1. Deriving a constructive model using an evolutionary simulation. (A)** Overview of our framework. An ensemble representing extant diversity is first generated then encoded as a sample-by-feature matrix and decomposed into discrete scales of entropy. The Layer-Restricted Boltzmann Machine (LRBM) approximates a generative process of extant diversity as a product of conditional distributions over scales of entropy where each scale is conditioned only the next-coarser one. **(B)** The simulation. Length-16 sequences are built under a set of hierarchical conditionally defined positional constraints (C1-C4), each governing a block of positions and conditioned on the preceding constraint (left). An ensemble of extant sequences is then produced by iterated cycles of mutation and selection on a fitness defined by those constraints (right). **(C)** Inference of entropic scales from the ensemble alone. Left: the per-position effective dimension (input) separates into four discrete clusters. These clusters recover the constraints shown in panel B. Right: the matrix of interaction weights in the resulting trained LRBM. Pixel strength from blue to red indicates interaction weight. **(D)** Comparison of the LRBM (top) with a profile model (each site is treated as an independent system component) (bottom). Left: distribution of Hamming distances of generated sequences from the natural ensemble through position 12. Middle: statistics of generated versus natural sequences for single-site, pairwise, and triplet correlations (Pearson *r*, inset). Right: distribution of fitness over generated sequences against the natural ensemble (grey distribution).

Pandey et al., Figure 2.

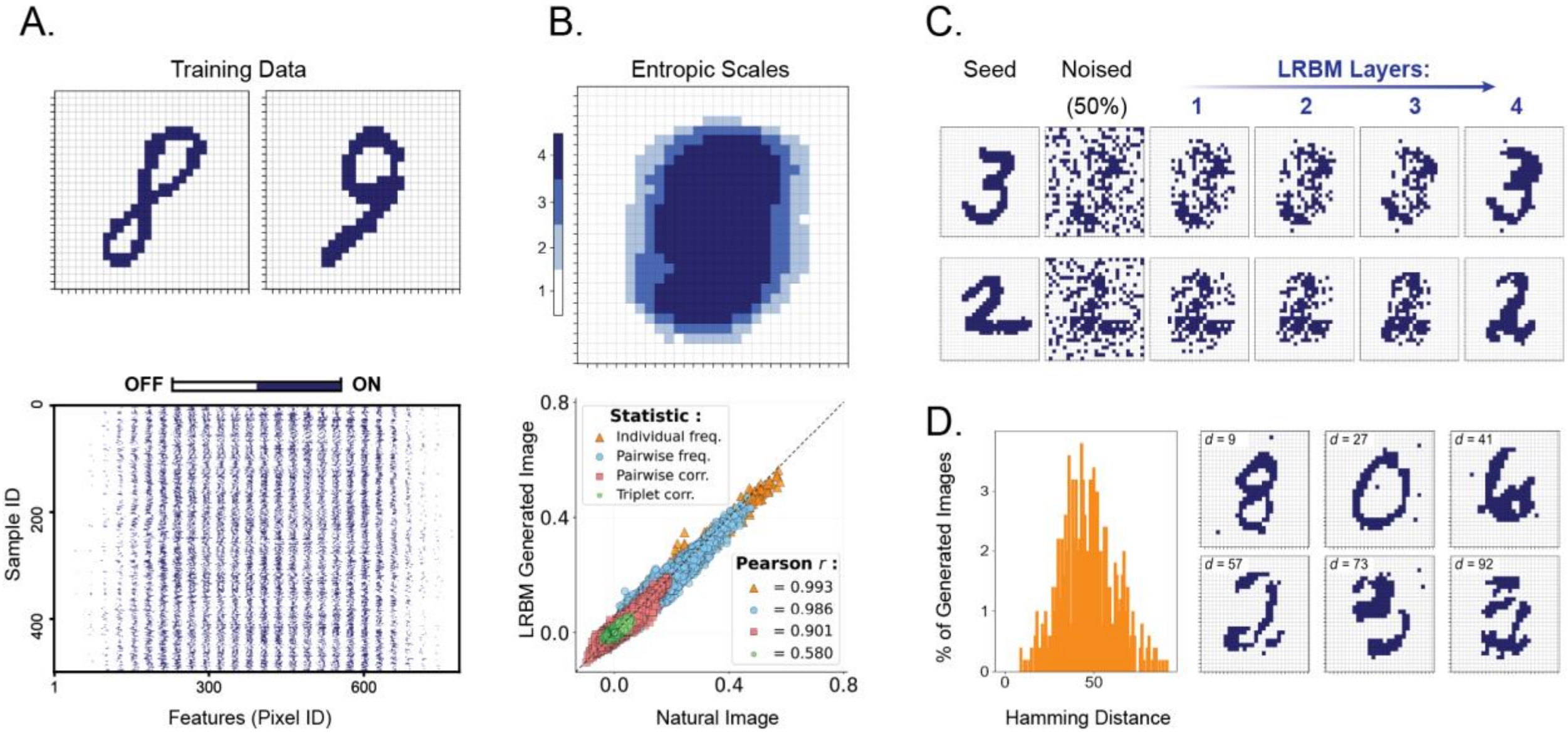

**Figure 2. A constructive model for handwritten digits. (A)** The ensemble comprises binarized 28 x 28 images of handwritten digits. Two examples are shown here (top). Each image is flattened to a binary feature vector in which every pixel is either OFF (white) or ON (blue), producing the sample-by-feature matrix used for training the LRBM (bottom). No labels of digit identity are included in the input data or used during the training process. **(B)** Top: the entropic scales of organization inferred from the ensemble with pixels colored according to their derived layer. Bottom: statistics of LRBM-generated images versus the training data. **(C)** Starting from a seed digit corrupted with 50% noise, the image is reconstructed layer by layer through the trained MNIST-LRBM. Two representative trajectories shown. **(D)** Distribution of Hamming distances between generated images and the nearest training example (left). Representative images generated from the MNIST-LRBM using pure noise as the input; Hamming distance (*'d'*) to nearest training data specified in the top-left corner (right).

Pandey et al., Figure 3.

A.

***Tabby cat is usually striped.***

***Cats are very clean animals.***

***Dog is a social animal.***

***Dogs can also be feral.***

***Bobcats are mostly nocturnal.***

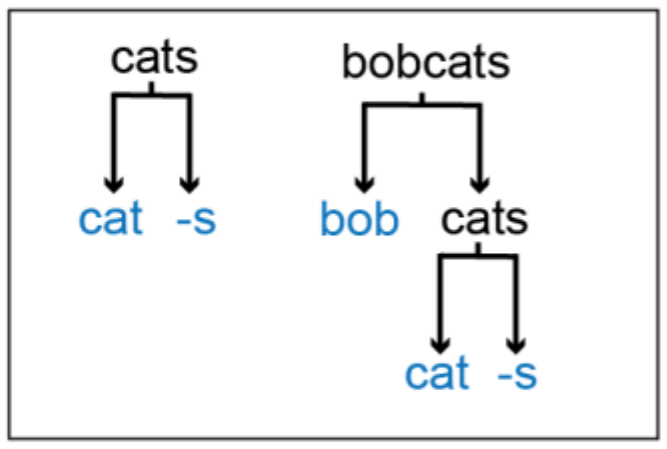

| tabby | cat | is | usually | striped | _ | _ |
|---|---|---|---|---|---|---|
| cat | -s | are | very | clean | animal | -s |
| dog | is | a | social | animal | _ | _ |
| dog | -s | can | also | be | feral | _ |
| bob | cat | -s | are | mostly | nocturnal | _ |

B.

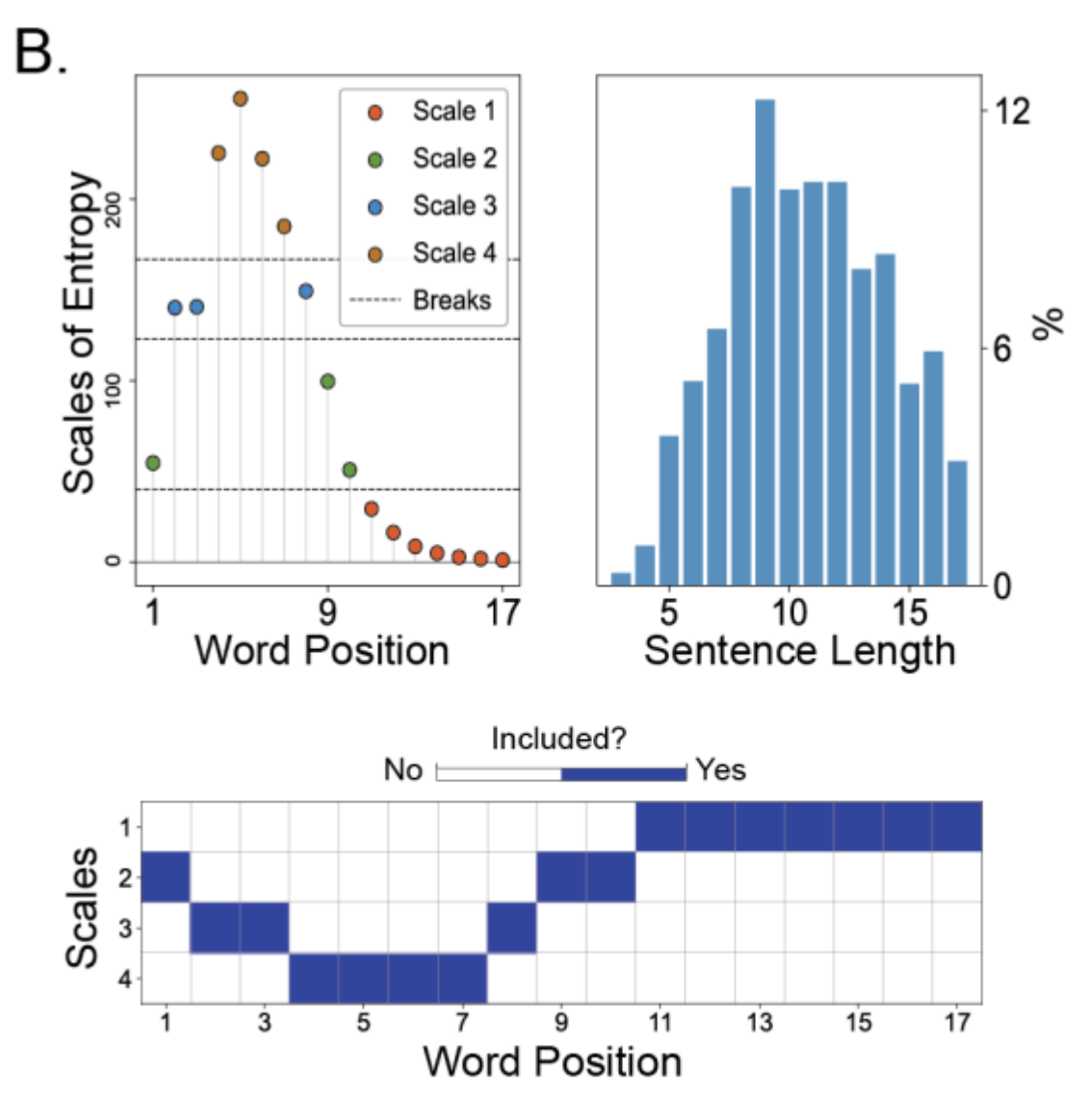

C.

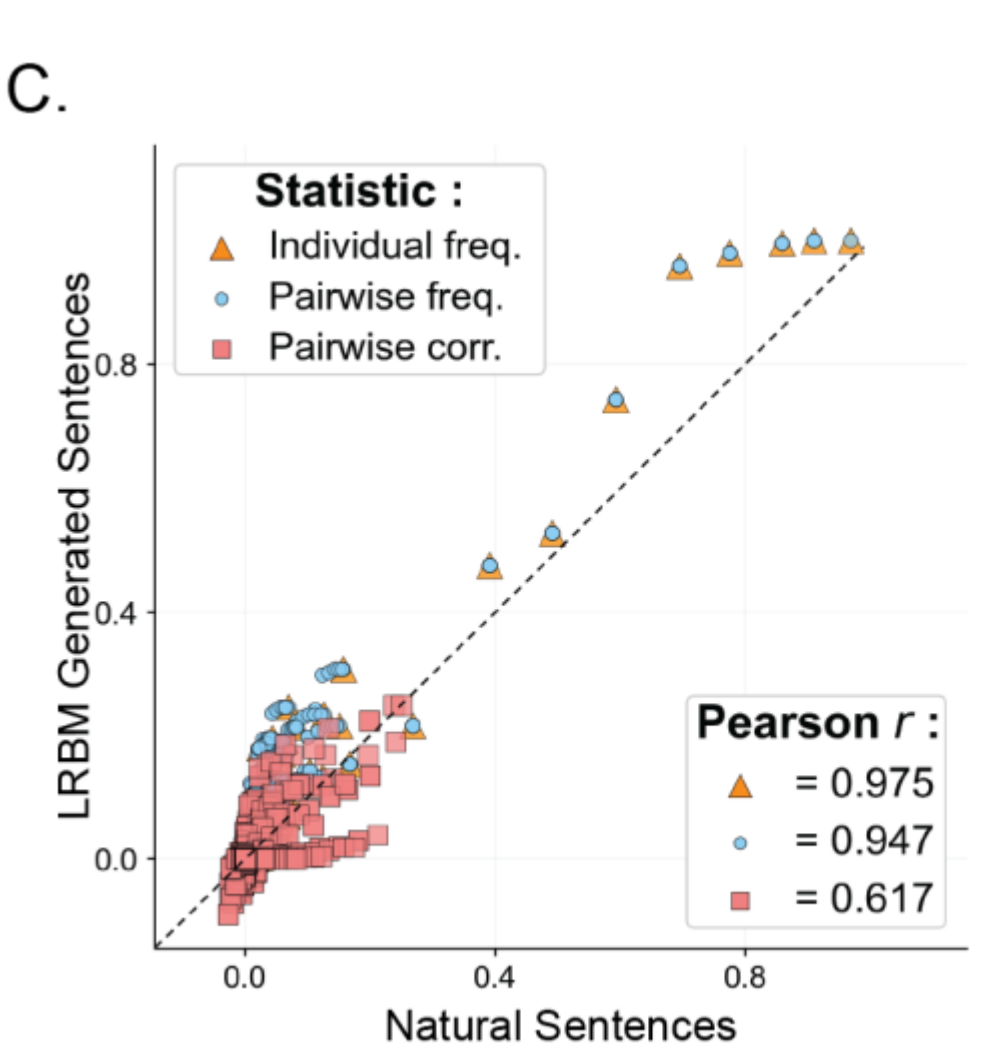

D.

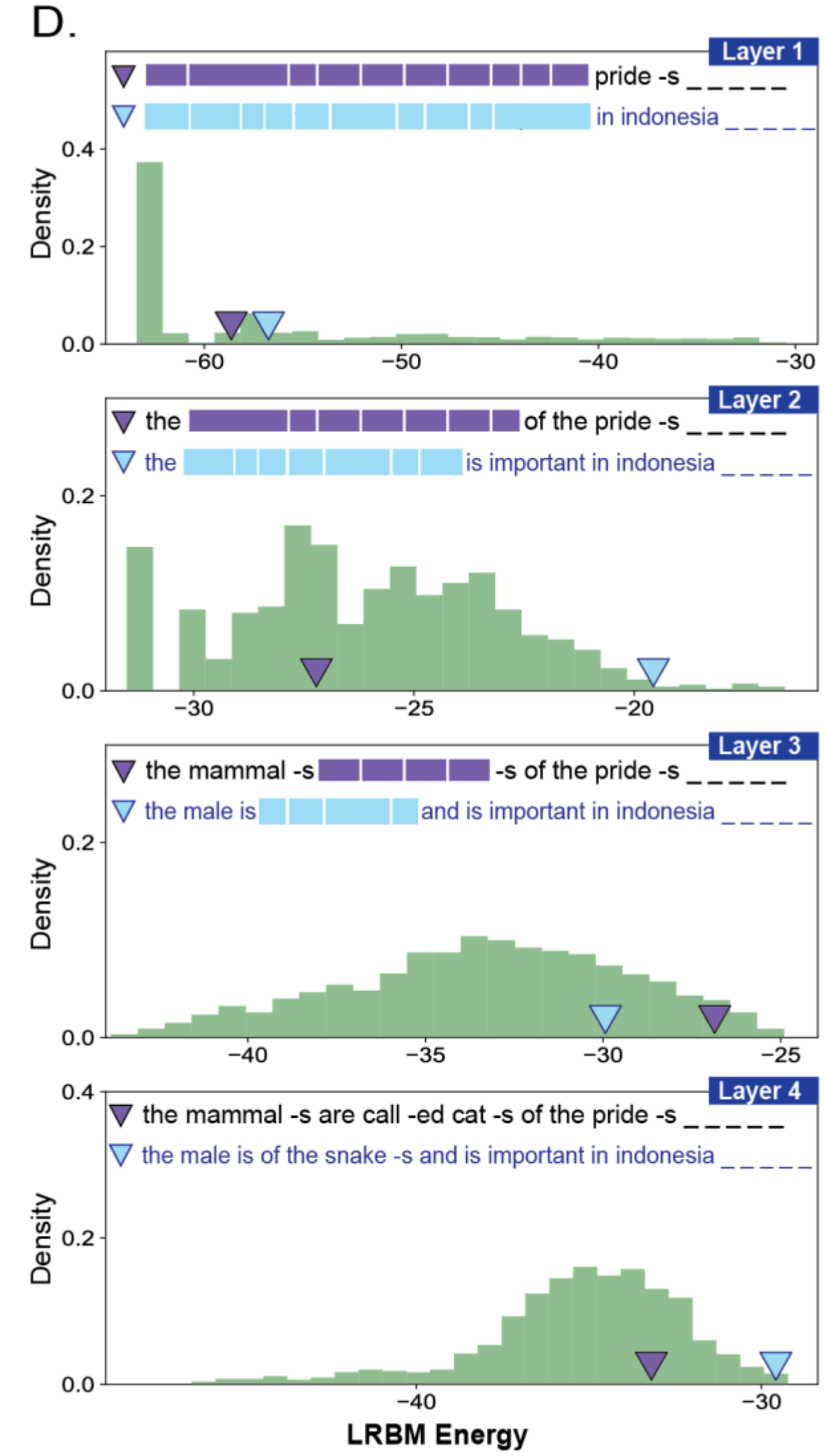

**Figure 3. Constructive generation of natural language sentences. (A)** Corpus preparation. Sentences from *'Simple Wikipedia'* articles on various animals were segmented into tokens, decomposing compound forms into their constituent morphemes (see **Methods**). Segmented sentences were aligned by word position into a fixed-length matrix of 16 positions with unoccupied terminal positions treated as gaps (right). **(B)** Entropic decomposition of the aligned corpus. Top left: effective dimension at each word where position is colored by the entropic cluster to which the position was assigned. Dashed lines indicate entropic breaks. Top right: distribution of sentence lengths across the corpus. Bottom: distribution of word positions (columns) belonging to entropic clusters (rows). **(C)** Summary statistics of SA-LRBM generated sentences plotted against the corresponding statistics of natural sentences: individual token frequencies, pairwise token frequencies, and pairwise correlations amongst tokens. The dashed line indicates identity; Pearson correlation coefficients provided for each statistic. **(D)** Synthetic sentence construction across SA-LRBM layers. Each panel shows the distribution of LRBM energies for generated sentences after satisfying Layers 1 (top-most panel) through 4 (bottom-most panel). Two representative constructive trajectories are shown above each distribution; colored blocks mark positions not yet assigned at the layer and tokens are displayed once determined. Triangles mark the energies of the two trajectories at the specific layer.

Pandey et al., Figure 4.

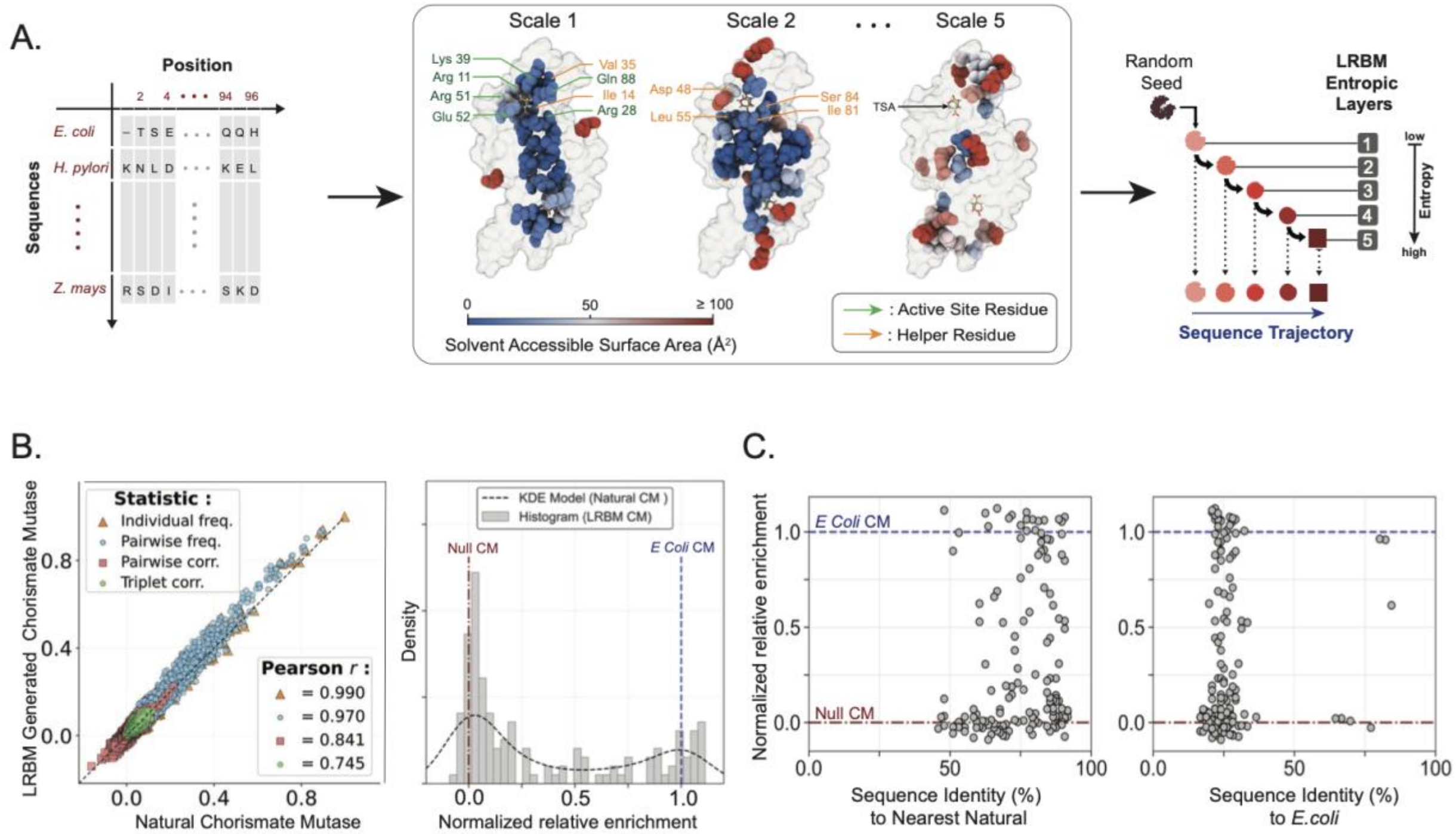

**Figure 4. LRBM-based generation of synthetic chorismate mutase enzymes. (A)** Construction of the chorismate mutase LRBM (CM-LRBM). Left: multiple sequence alignment of natural chorismate mutase homologs. Middle: positions assigned to entropic scales 1, 2, and 5 mapped onto the structure of *E. coli* CM (PDB: 1ECM) and rendered as spheres colored by solvent-accessible surface area. The bound transition-state analog is shown in stick representation. Green arrows mark active site residues ('Active Site Residue') and orange arrows mark residues around the active site known to effect catalytic activity ('Helper Residue'). Right: schematic of LRBM-based CM generation. CM-LRBM comprised five layers; a random seed sequence is progressed through all five layers resulting in an LRBM-generated sequence. **(B)** Left: summary statistics of 2,000 LRBM-generated CM sequences plotted against the corresponding statistics of natural CM sequences within the MSA in panel A. Dashed line indicates identity; Pearson correlation coefficients are shown for each statistic. Right: distribution of normalized relative enrichment for LRBM-generated sequences assayed by *in vivo* complementation (histogram) overlaid with a kernel density estimate (KDE) of the relative enrichment distribution for all natural CM sequences present within the MSA (dashed line). Values were normalized such that 0 relative enrichment corresponds to the null (non-functional) control and 1.0 to *E. coli* CM. **(C)** Normalized relative enrichment of individual LRBM-generated sequence plotted against sequence identity to the nearest natural CM (left) and to *E. coli* CM (right). Dashed lines indicate the *E. coli* CM and the null CM as in panel B.

Pandey et al., Figure 5.

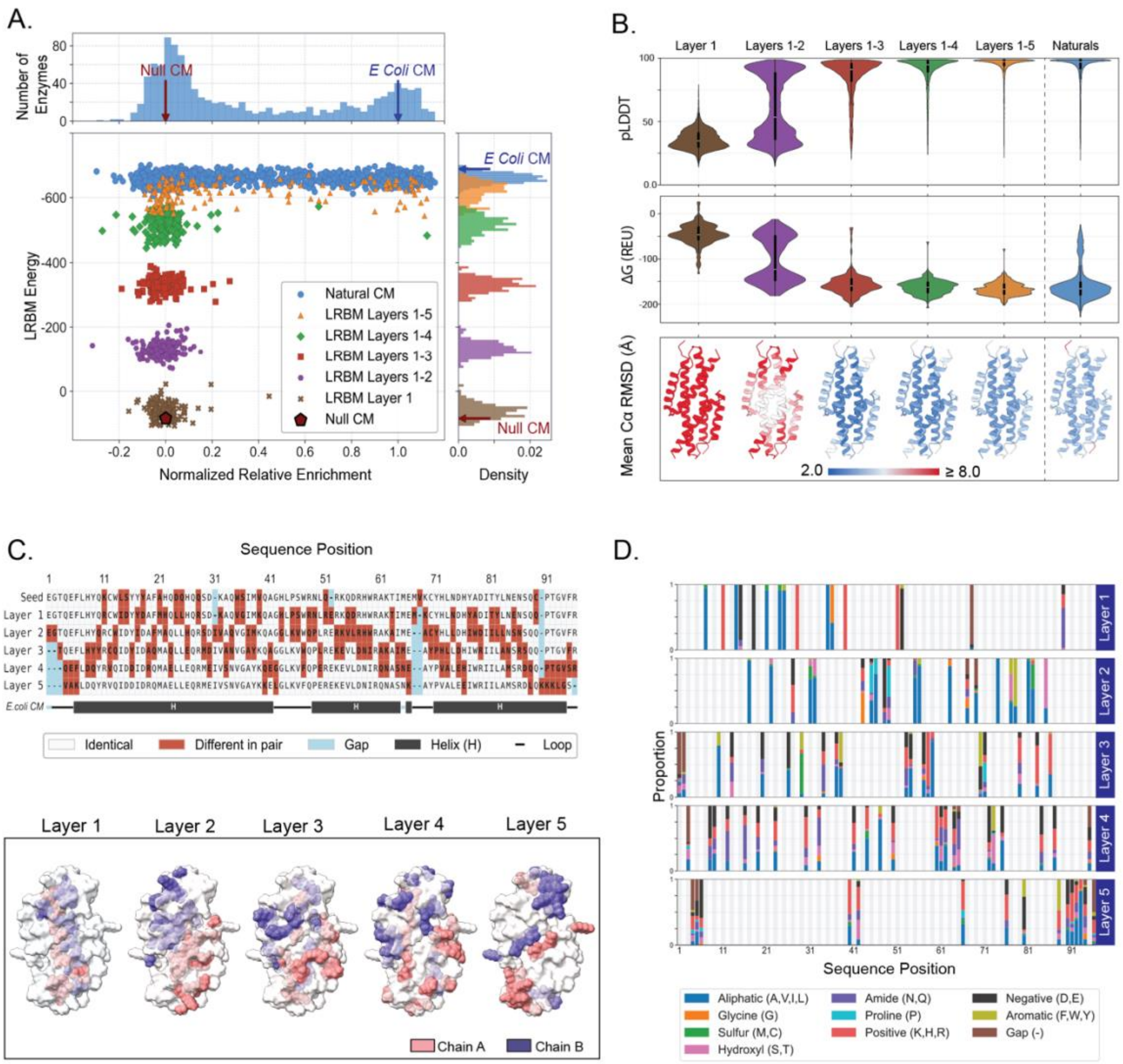

**Figure 5. Layer-wise progression of energy, structure, and composition in constructive CM design. (A)** Top: Distribution of normalized relative enrichment for all natural CMs in MSA (see **Fig. 4A**). Normalized relative enrichment plotted against LRBM energy for natural CM sequences (blue dots), the null CM sequence (red pentagon), and designed sequences generated by satisfying CM-LRBM layer 1 through layer 5 (colored dots). **(B)** Structural evaluation of constructed sequences from panel A and natural CMs in MSA by AlphaFold3 (top) and Rosetta (middle). Bottom panel shows overlay of predicted CM structures at each CM-LRBM layer colored by root-mean squared deviation of the backbone alpha carbon atoms. **(C)** Single constructive trajectory. Top: alignment of the random seed sequence and intermediate sequences produced after each CM-LRBM layer. Positions differing from the preceding sequence in the pair are displayed in red and gaps in blue. Secondary structure of *E.coli* CM displayed below alignment. Bottom: Corresponding predicted structures with positions colored as they are being defined in each CM-LRBM layer. **(D)** Positional amino acid composition of designed sequences that function like *E. coli* CM by CM-LRBM layer. Positions defined in each layer are colored by physicochemical characteristics (see color key).

Pandey et al., Figure 6.

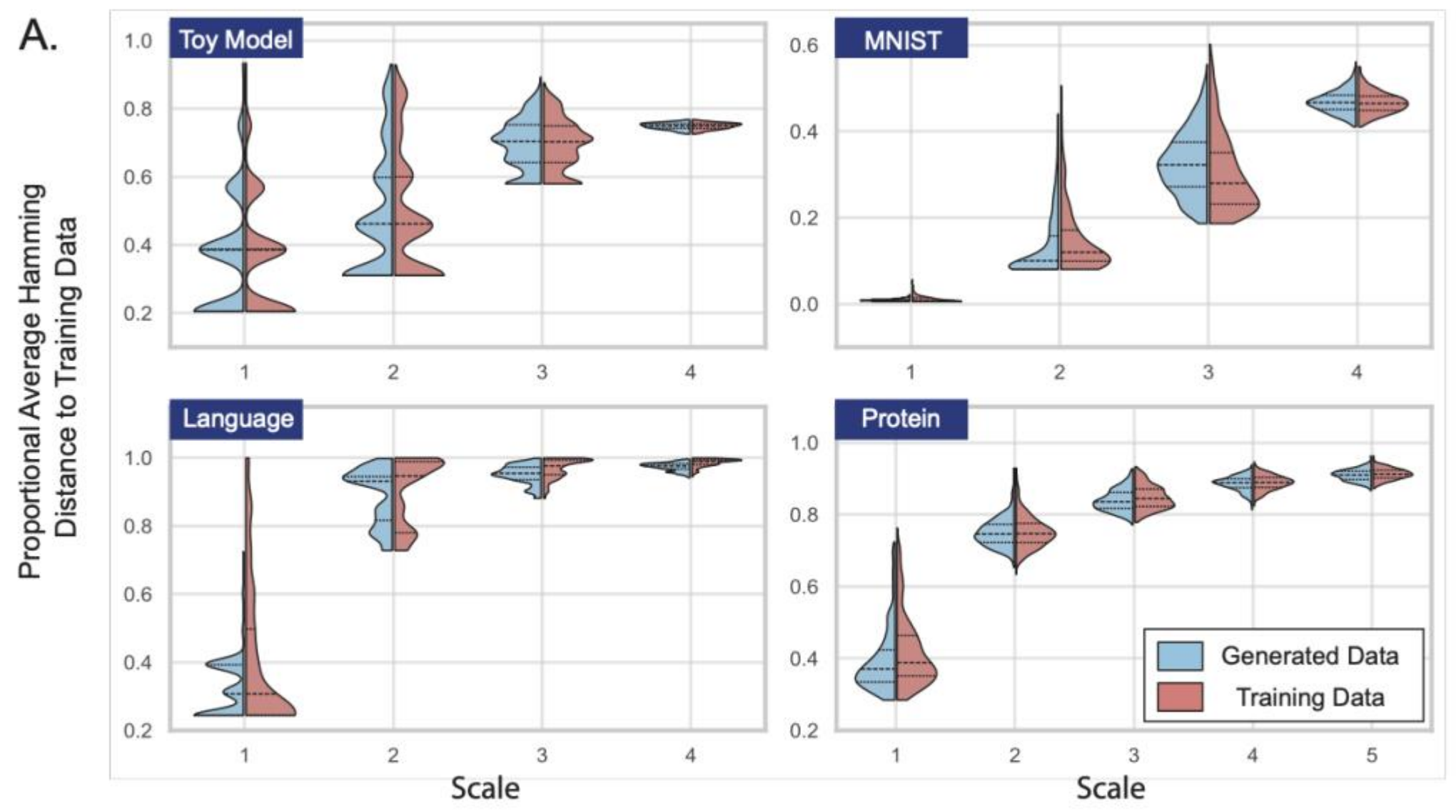

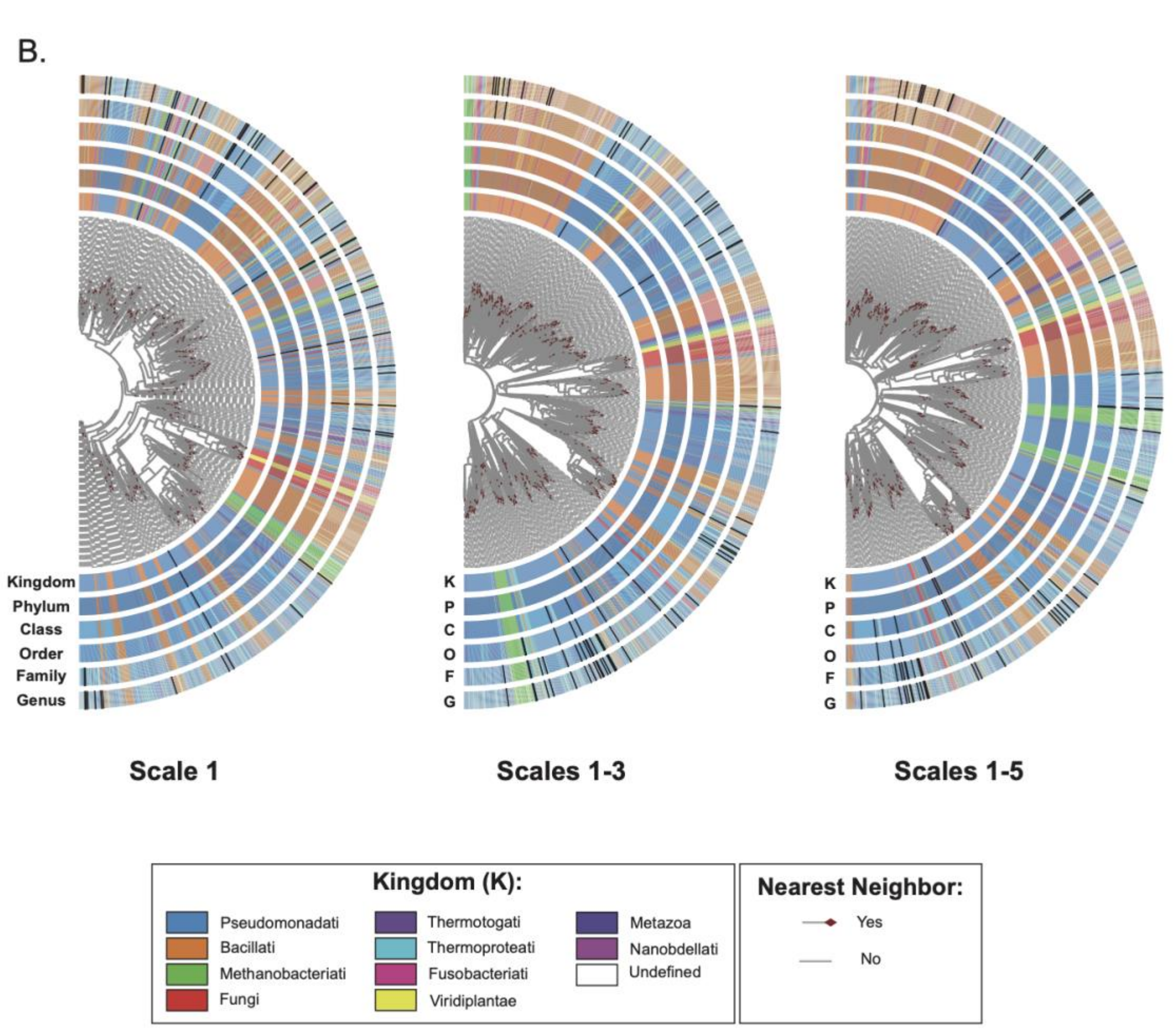

**Figure 6. Origins of novelty in LRBM models. (A)** Proportional average Hamming distance of LRBM-generated systems relative to the data used to train LRBM models, computed separately over system components belonging to each entropic scale. Blue distributions reflect distances of LRBM-generated samples to the training data; red distributions reflect distances of training data to itself. Distances normalized by number of positions in each scale. Across domains, distances increase with scale; generated and training distributions overlap. **(B)** Phylogenetic trees of natural CM sequences constructed from considering positions reflected by entropic scale 1 (left), up to scale 3 (middle), and up to scale 5 (right). Phylogenetic designation at scale of Kingdom shown in color key and displayed in each tree. Nearest neighbor of CM-LRBM generated synthetic sequences shown in red dots.

| System | Training Data Size | Number of Position | Total Alphabet Size | #LRBM Parameter ($N_{LRBM}$) | Model Size (MB) |
|---|---|---|---|---|---|
| MNIST LRBM | 500 | 784 | 2 | 855,346 | 29.3 |
| SA-LRBM (*'Simple Wikipedia'* on Animals) | 1589 | 17 | 2156 | 14,672,394 | 439.1 |
| AW-LRBM ('*Alice in Wonderland*') | 716 | 17 | 799 | 3,124,135 | 92.8 |
| CM-LRBM (Chorismate Mutase) | 1130 | 96 | 21 | 1,242,234 | 38.9 |

**Table 1. Training data and model dimensions for each LRBM.** Training data size indicates the number of samples used to fit each LRBM model; number of positions indicates the length of the aligned representation; total alphabet size indicates the number of states available at each position. Total number of LRBM model parameters ($N_{LRBM}$) and model size for all models reported in last two columns.

## Supplementary Figures

A.

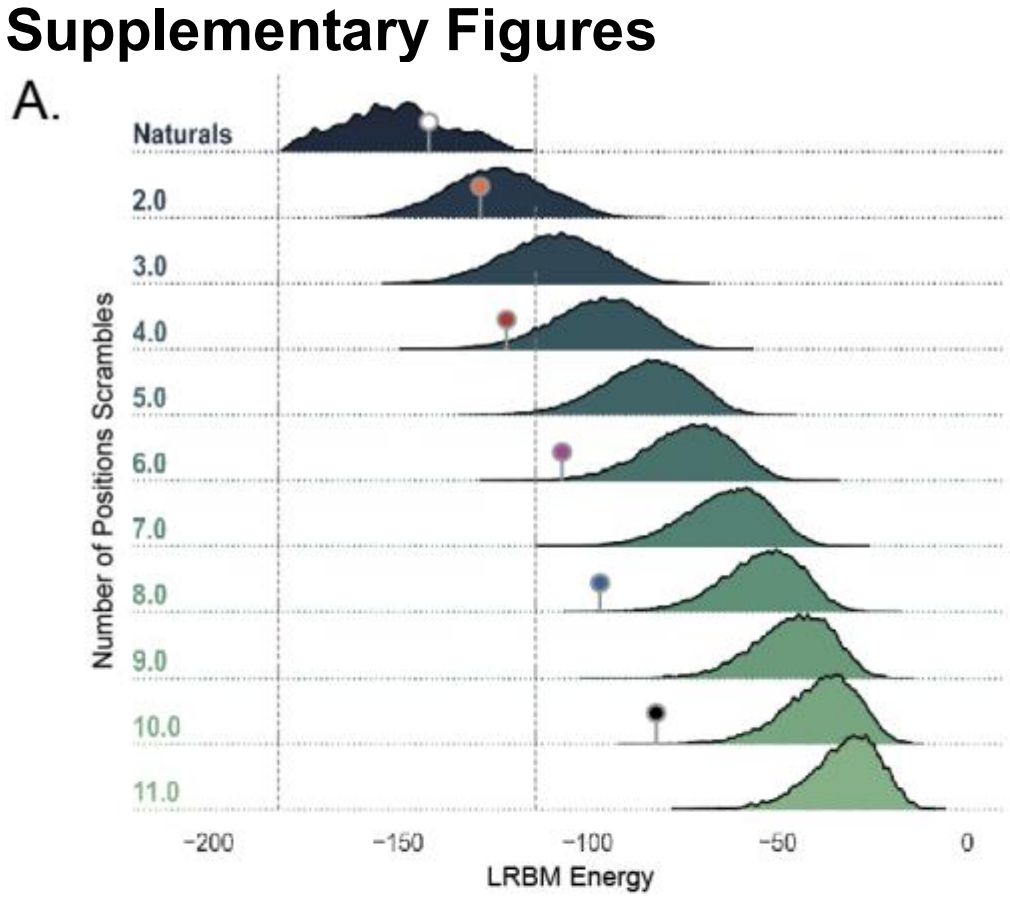

| Marker | #Positions Scrambled | Sentence | LRBM Energy |
|---|---|---|---|
| | 0 | this mean -s they are awake at night and a- sleep during the day _ _ _ | -142.82 |
| | 2 | this mean -s are they awake at night and a- sleep during the day _ _ _ | -129.21 |
| | 4 | this mean -s they are awake at night and a- the sleep day during _ _ _ | -122.27 |
| | 6 | this mean -s are they awake at night and a- the day during sleep _ _ _ | -107.79 |
| | 8 | this mean -s are sleep the day night and a- awake they at during _ _ _ | -97.76 |
| | 10 | this mean -s are sleep during the night day they at and awake -a _ _ _ | -82.97 |

B.

**Layer 4**

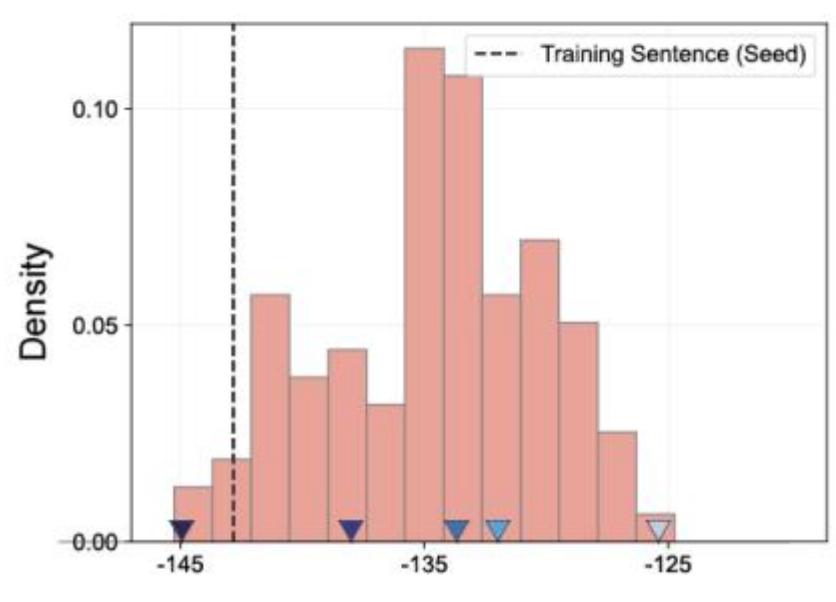

| Marker | Sentence |
|---|---|
| — — | this mean -s they are awake at night and a- sleep during the day _ _ _ |
| | this mean -s are call -ed the night and a- sleep during the day _ _ _ |
| | this mean -s are very dominant in night and a- sleep during the day _ _ _ |
| | this mean -s and breed -s in night and a- sleep during the day _ _ _ |
| | this mean -s use -d in the night and a- sleep during the day _ _ _ |
| | this mean -s are tame -r in night and a- sleep during the day _ _ _ |

**Layers 3 and 4**

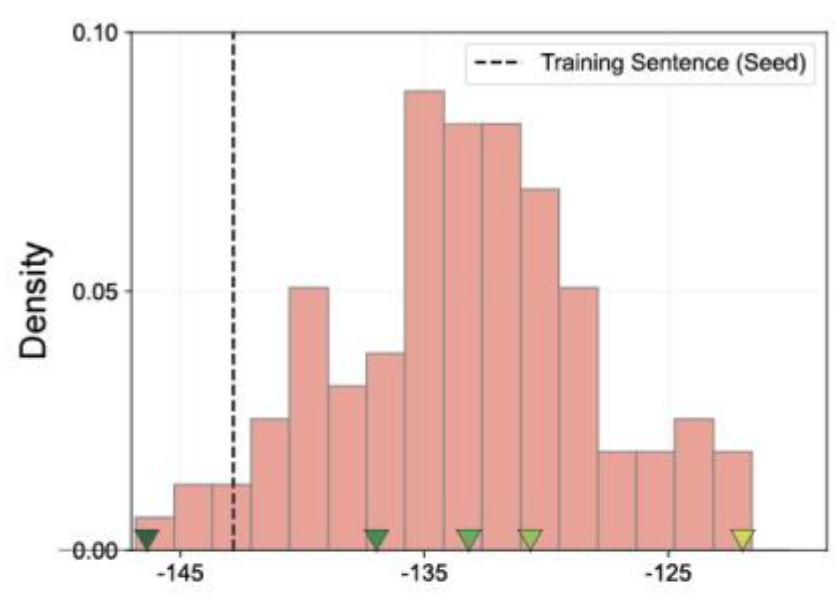

| Marker | Sentence |
|---|---|
| | this mean -s are call -ed the night and a- sleep during the day _ _ _ |
| | this is be- cause predator -s can not and a- sleep during the day _ _ _ |
| | this mammal -s are found in short -s and a- sleep during the day _ _ _ |
| | this is a predator -s of locusts -s and a- sleep during the day _ _ _ |
| | this is usually for a time with -s and a- sleep during the day _ _ _ |

**Layers 2 through 4**

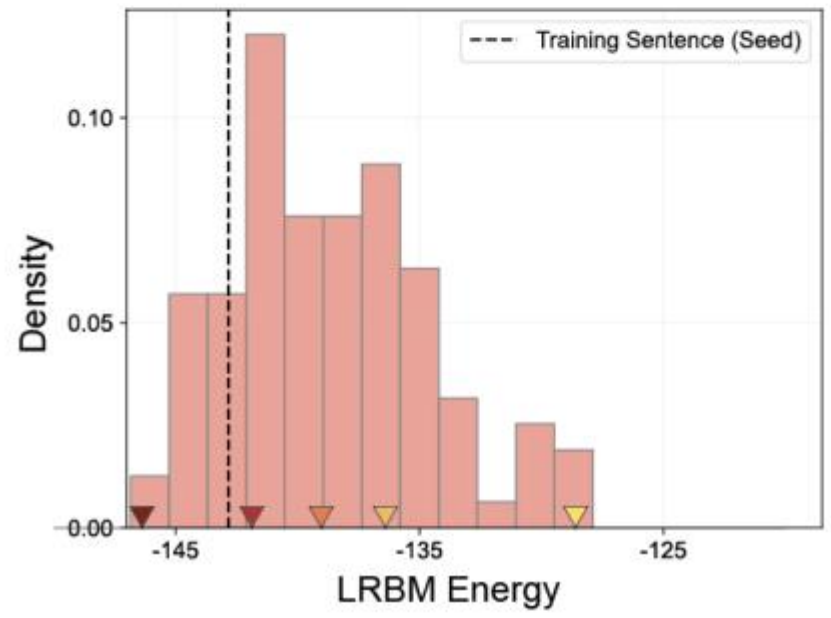

| Marker | Sentence |
|---|---|
| | there are about fifty different species of rabbit -s and sleep during the day _ _ _ |
| | pig -s are often call -ed the mammal -s and sleep during the day _ _ _ |
| | the lion -ess will move the cub -s and a sleep during the day _ _ _ |
| | bear -s are know -n for being mammal -s of sleep during the day _ _ _ |
| | the male cat keep the deer population -ed to a sleep during the day _ _ _ |

**Figure S1. Information content of the SA-LRBM model. (A)** (Left). Histogram of SA-LRBM energies (x-axis) for (i) '*Simple Wikipedia'* corpus sentences ('Naturals') and (ii) sentences with two to eleven words scrambled (y-axis). (Right). Example of a natural sentence scrambled with associated energies; markers correspond to markers displayed on histograms in left panel. **(B)** (Left). Distribution of LRBM energies (x-axis) for 100 unique sentences generated by melting SA-LRBM Layer 4 (top), Layers 3 and 4 (middle), and Layers 2 through 4 (bottom) and then reconstructing sentences. (Right) Example sentences; markers correspond to markers displayed on histograms in left panel. Word positions in red are melted and reconstructed while positions in black are static.

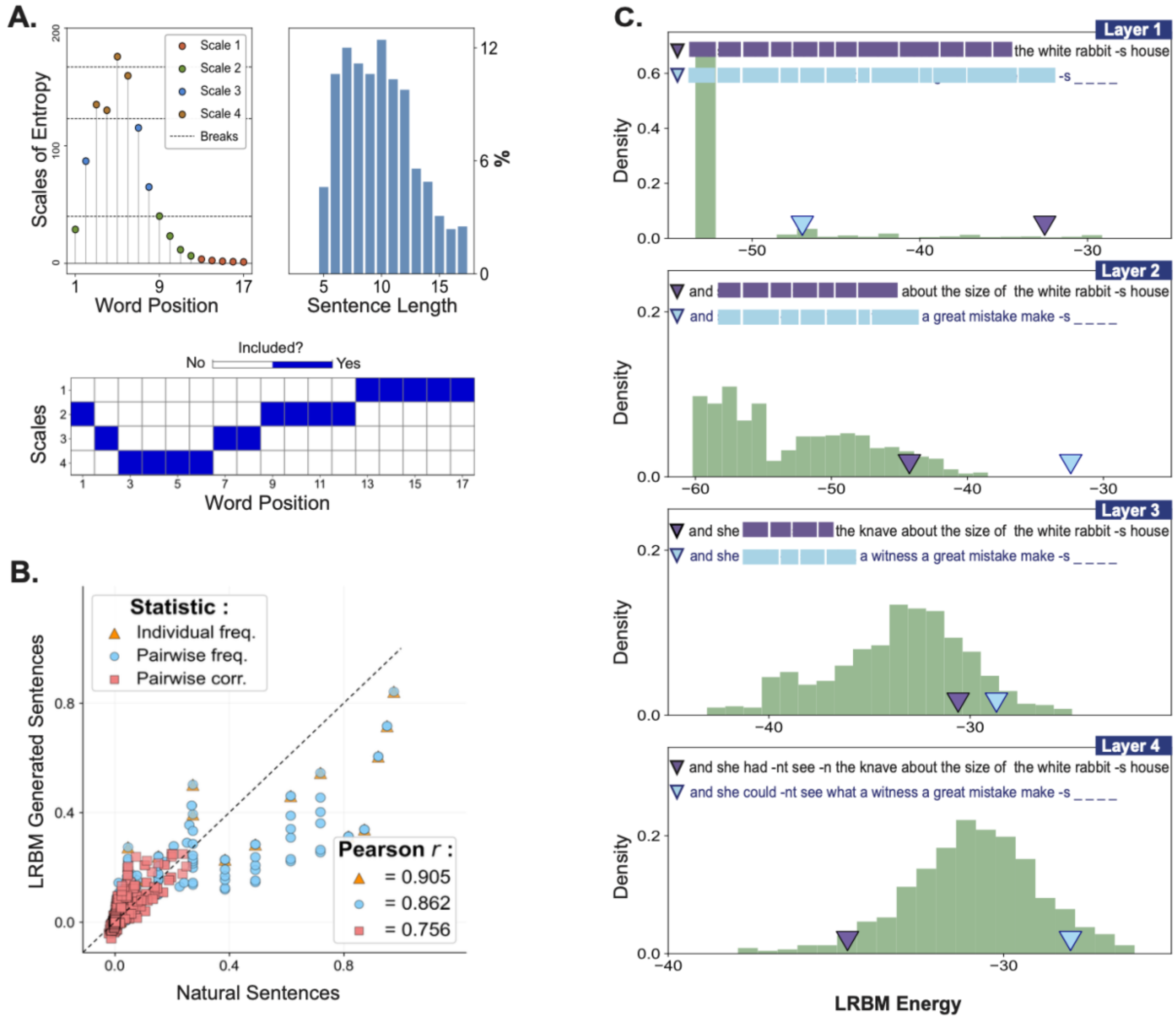

**Figure S2. Constructive generation of sentences from *Alice in Wonderland*. (A)** Entropic decomposition of the aligned corpus. Top left: scale of entropy at each word where position is colored by the entropic cluster to which the position was assigned. Dashed lines define scales of entropy. Top right: distribution of sentence lengths across the input corpus. Bottom: distribution of word positions (columns) belonging to entropic clusters (rows). **(B)** Summary statistical of AW-LRBM generated sentences plotted against the corresponding statistics of natural sentences: individual token frequencies, pairwise token frequencies, and pairwise correlations amongst tokens. Dashed line indicates identity; Pearson correlation coefficients provided for each statistic. **(C)** Synthetic sentence construction across SA-LRBM layers. Each panel shows the distribution of LRBM energies (x-axis) for generated sentences after satisfying Layers 1 (top-most panel) through 4 (bottom-most panel). Two representative constructive trajectories are shown above each distribution; colored blocks within constructive trajectories mark positions within the sentence not yet assigned at the layer. Triangles mark the energies of the two trajectories at the specific layer.

A.

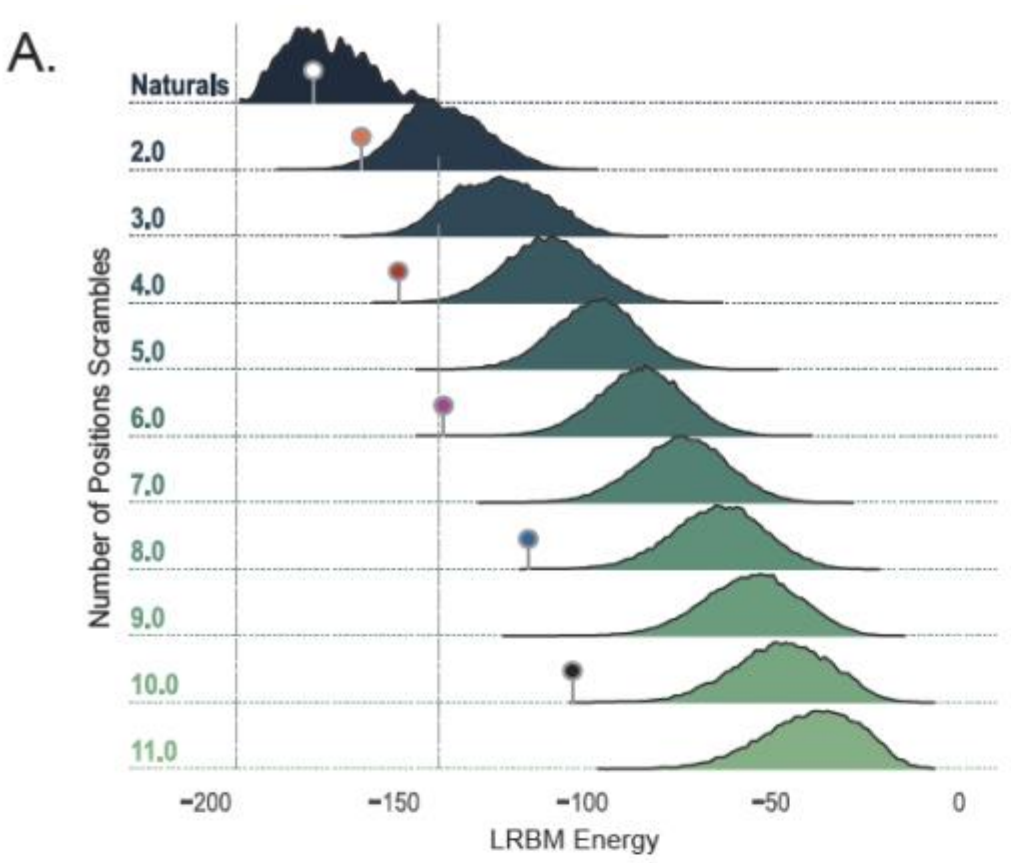

| Marker | #Positions Scrambled | Sentence | LRBM Energy |
|---|---|---|---|
| | 0 | well you see it had to go and visit the duchess _ _ _ _ _ _ | -172.27 |
| | 2 | well you see it had and go to visit the duchess _ _ _ _ _ _ | -159.39 |
| | 4 | well you it to see had go and visit the duchess _ _ _ _ _ _ | -149.88 |
| | 6 | well you see it had to go and _ _ _ _ _ _ the duchess visit | -137.38 |
| | 8 | to you see it had duchess go and _ _ _ _ _ _ well the visit | -115.13 |
| | 10 | well you it visit duchess had go and _ _ _ _ _ _ the see to | -103.30 |

B.

**Layer 4**

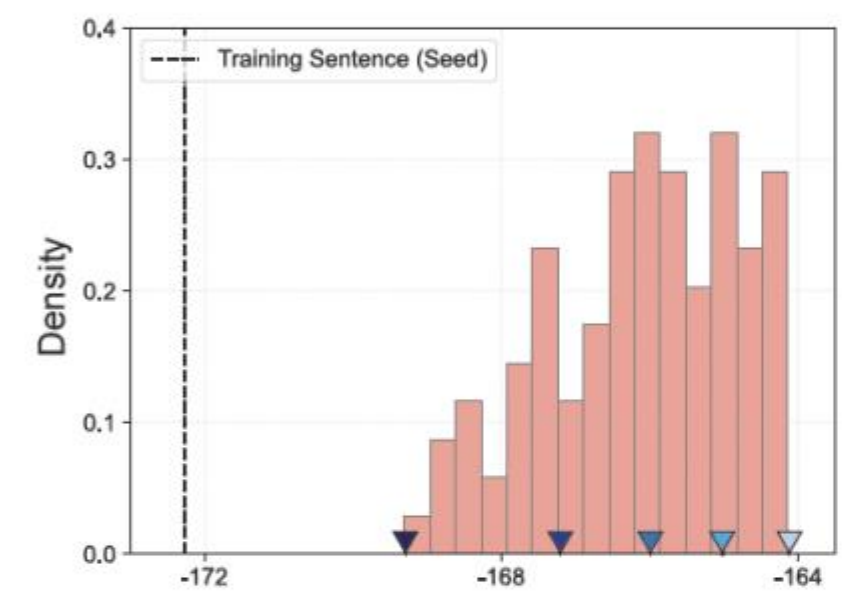

| Marker | Sentence |
|---|---|
| | well you see it had to go and visit the duchess_ _ _ _ _ _ |
| | well you like to hear what go and visit the duchess_ _ _ _ _ _ |
| | well you have to roll about go and visit the duchess_ _ _ _ _ _ |
| | well you could -nt see very go and visit the duchess_ _ _ _ _ _ |
| | well you see how hes trembling go and visit the duchess_ _ _ _ _ _ |
| | well you knew you would -nt go and visit the duchess_ _ _ _ _ _ |

**Layers 3 and 4**

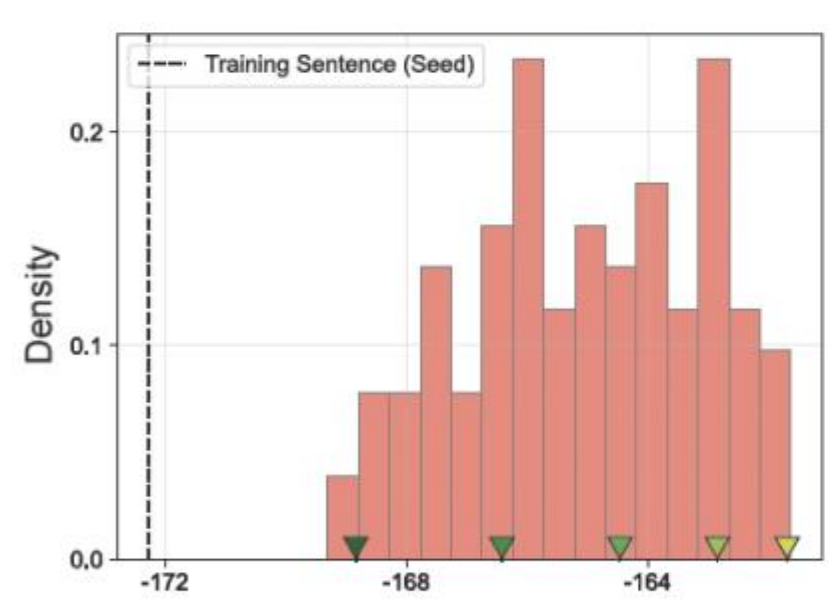

| Marker | Sentence |
|---|---|
| | well you like to know what go and visit the duchess_ _ _ _ _ _ |
| | well you was very rude to go and visit the duchess_ _ _ _ _ _ |
| | well she was -nt a pig that great visit the duchess_ _ _ _ _ _ |
| | well alice had -nt see -n the queen visit the duchess_ _ _ _ _ _ |
| | well she had a pair of the world visit the duchess_ _ _ _ _ _ |

**Layers 2 through 4**

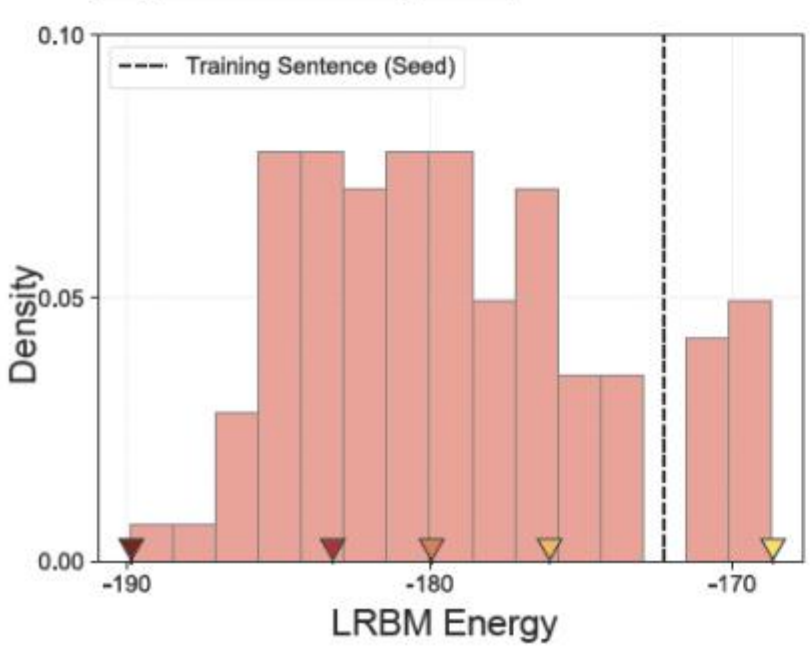

| Marker | Sentence |
|---|---|
| | when she said to herself_ _ _ _ _ _ _ _ _ _ _ _ _ |
| | when she look at the picture _ _ _ _ _ _ _ _ _ _ _ |
| | and he thought she was his housemaid _ _ _ _ _ _ _ _ _ _ |
| | how ever she soon came to a great mushroom _ _ _ _ _ _ _ _ |
| | you know lucky she found the quantities of leg -s _ _ _ _ _ _ _ |

**Figure S3. Information content of the AW-LRBM model.** (Left). Histogram of AW-LRBM energies (x-axis) for (i) '*Alice in Wonderland'* corpus sentences ('Naturals') and (ii) sentences with two to eleven words scrambled (y-axis). (Right). Example of a natural sentence scrambled with associated energies; markers correspond to markers displayed on histograms in left panel. **(B)** (Left). Distribution of LRBM energies (x-axis) for 100 unique sentences generated by melting AW-LRBM Layer 4 (top), Layers 3 and 4 (middle), and Layers 2 through 4 (bottom) and then reconstructing sentences. (Right) Example sentences; markers correspond to markers displayed on histograms in left panel. Word positions in red are melted and reconstructed while positions in black are static.

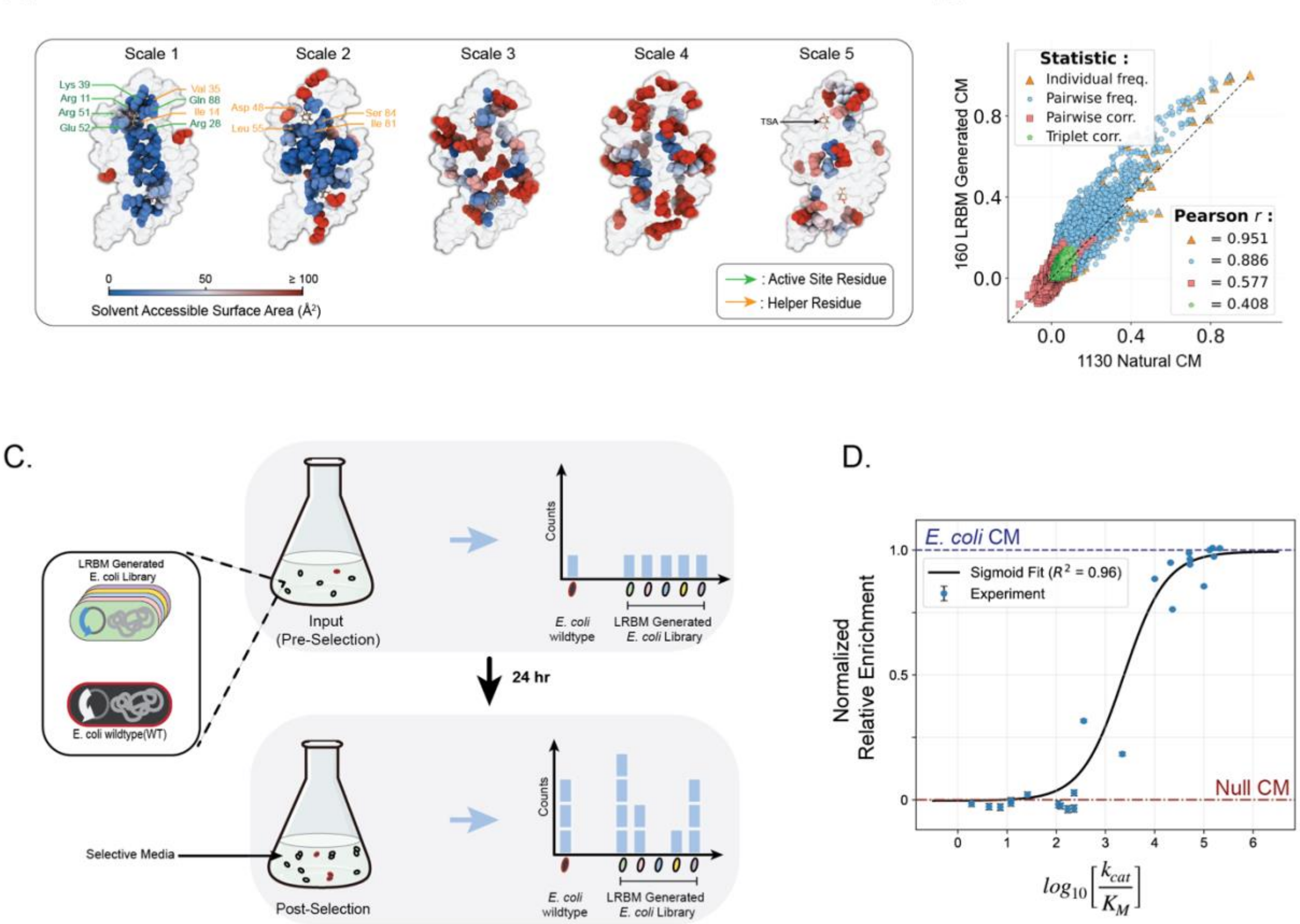

**Figure S4. (A)** Residues constituting entropic scales 1 through 5 mapped onto the structure of *E. coli* CM (PDB: 1ECM) and rendered as spheres colored by solvent-accessible surface area. The bound transition-state analog is shown in stick representation. Green arrows mark active site residues ('Active Site Residue') and orange arrows mark residues around the active site known to effect catalytic activity ('Helper Residue'). **(B)** Summary statistics of 160 randomly chosen CM-LRBM generated sequences plotted against the corresponding statistics of natural sequences where the dashed line indicates identity. **(C)** Chorismate Mutase Assay: Both synthetic and wildtype sequences are co-cultured in selective media condition for 24 hours. Pre- and post-selection sequencing counts were used to determine relative enrichment **(D)** Points represent single mutants of CM. Normalized relative enrichment of the mutant library (y-axis) as a function of their catalytic activity (x-axis) is plotted. Black line is the sigmoid curve fit for this relation.

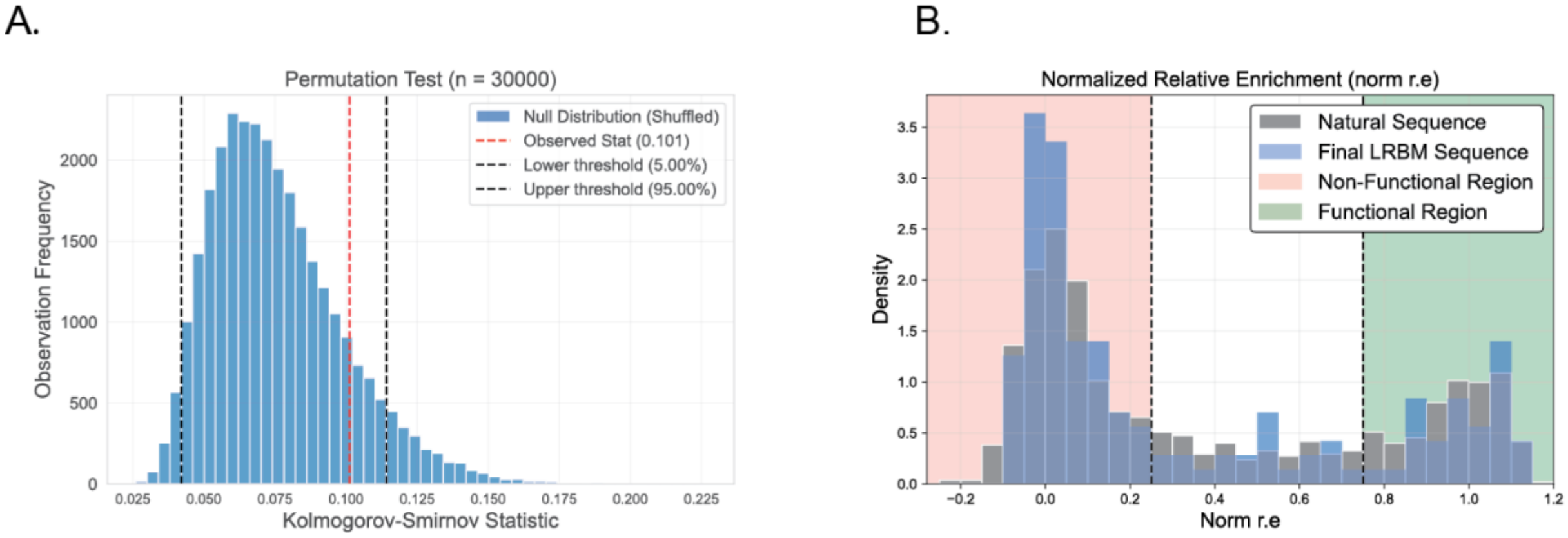

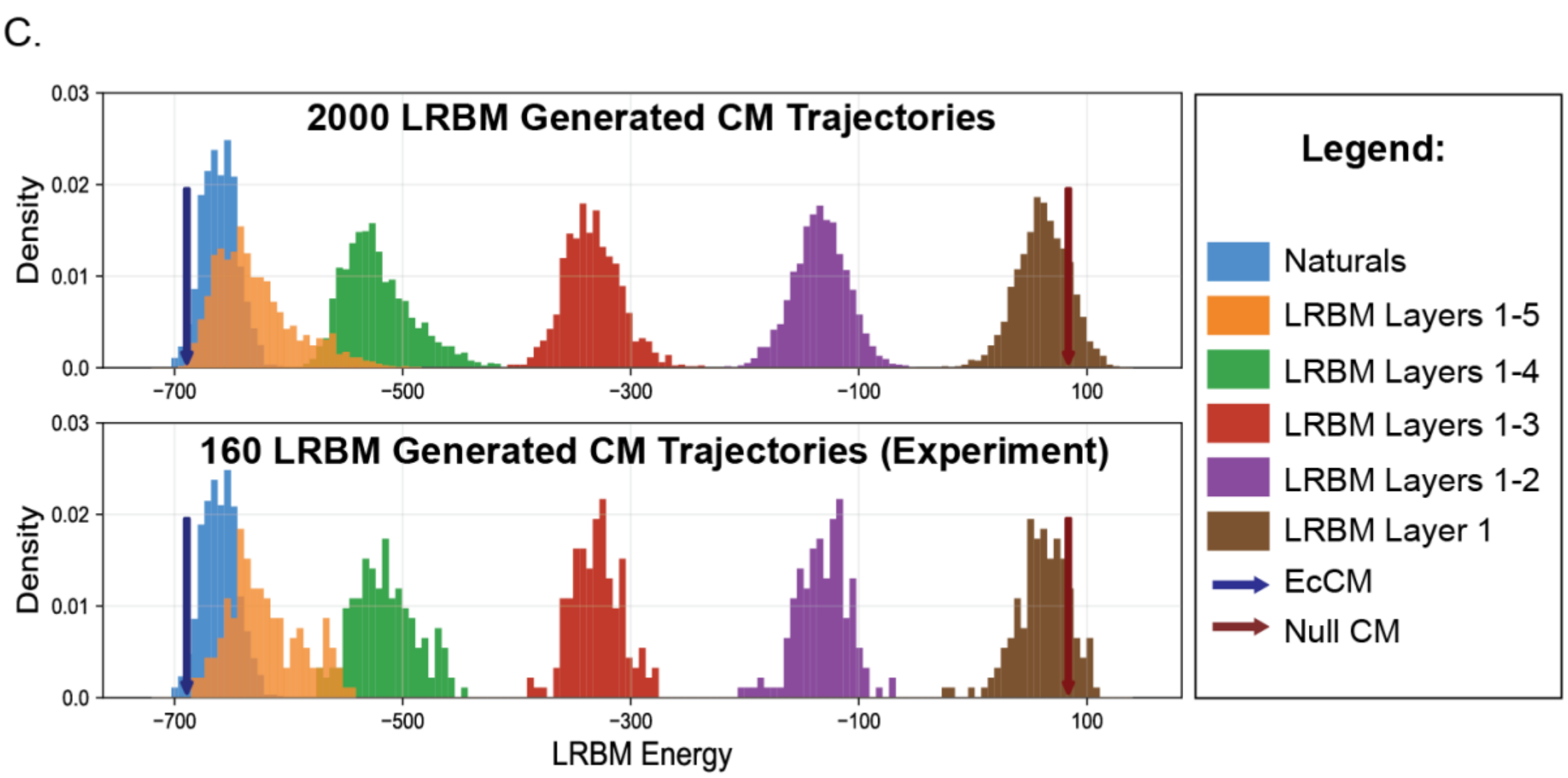

**Figure S5. (A)** A 30,000-trial Kolmogorov-Smirnov permutation test showing that the distribution of 160 CM-LRBM final sequences is statistically indistinguishable from 1130 natural sequences, with the observes statistic falling well within 5%-95% null distribution. **(B)** Delineation of functional and non-functional sequences as judged by normalized relative enrichment (Norm r.e., x-axis). Norm r.e. $\geq$ 0.75 (green shaded area) indicates functional sequences; Norm r.e $\leq$ 0.25 (red shaded area) indicates non-functional sequences**.** Included in the histogram are both natural and CM-LRBM designed sequences. **(C)** Layer wise CM-LRBM energies for the 2000 CM-LRBM designed sequences and the subset of 160 sequences used in experiments.

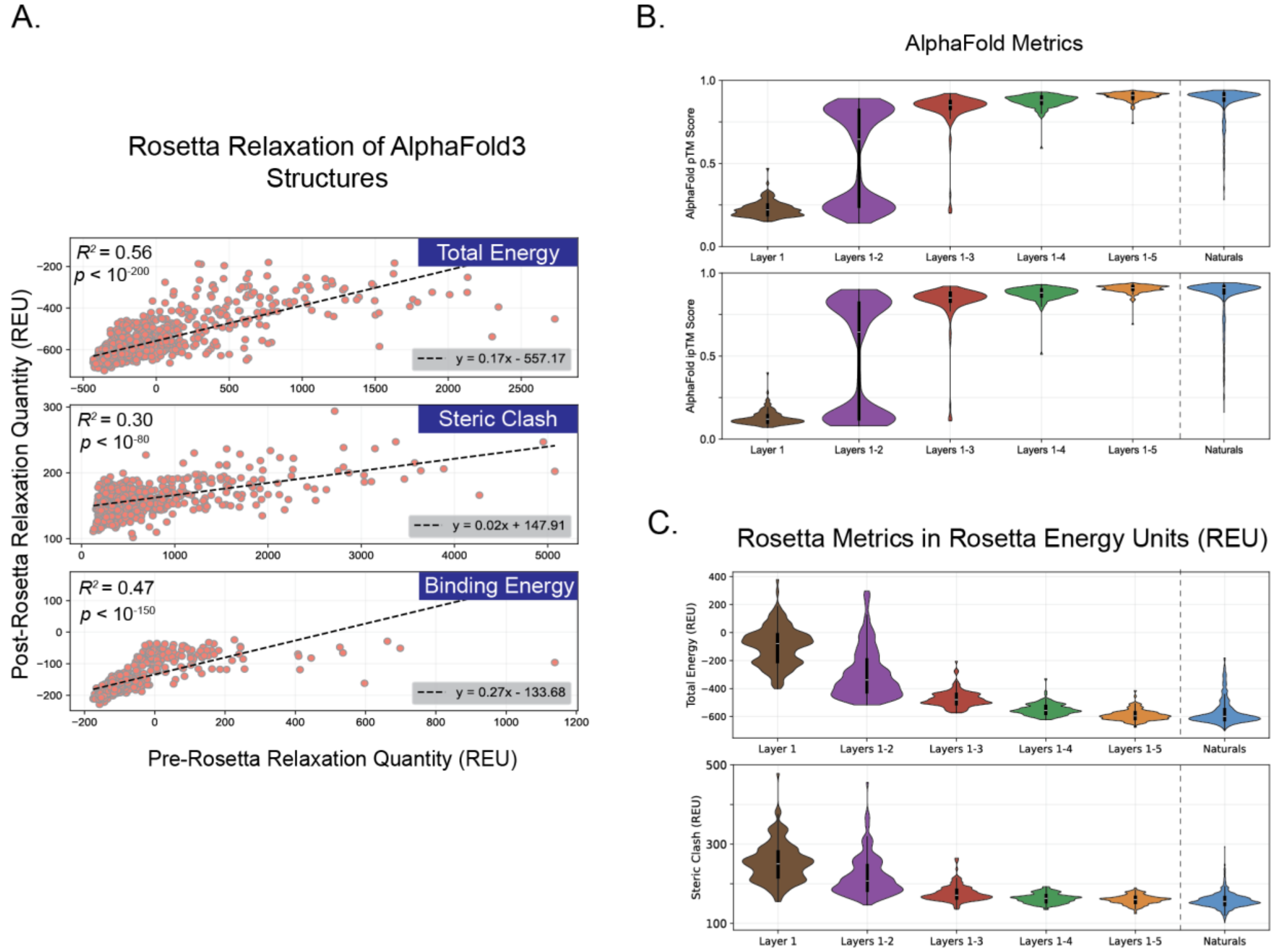

**Figure S6. (A)** Relationship between AlphaFold and Rosetta post-relaxation energy metrics with individual points representing individual AlphaFold structures. Rosetta relaxation dramatically reduces these metrics in absolute value while preserving the rank-order revealing a consensus between the two models regarding the stability of a given structure. **(B,C)** Metrics for structural evaluation of constructed sequences trajectories and natural CMs in MSA by AlphaFold3 (panel B; pTM score reflective of overall protein structure and ipTM score reflective of interface stability) and Rosetta (panel C; Total Energy and Steric Clash) also reveal that structure stabilization is encoded after layer 3.

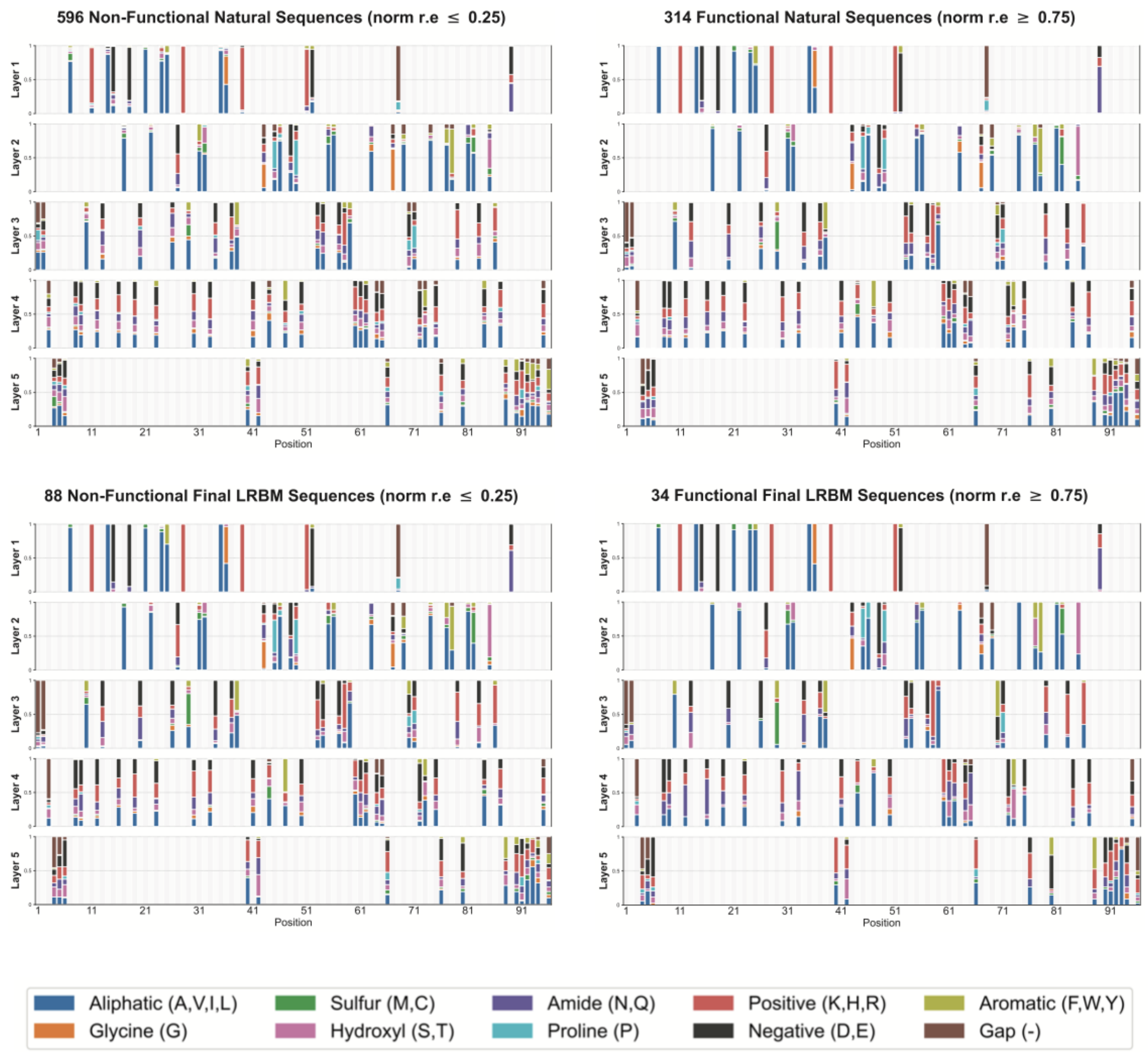

**Figure S7.** Positional amino acid composition of functional and non-function natural and CM-LRBM designed sequences by CM-LRBM layer. Positions defined in each layer are colored by physicochemical characteristics (see color key).

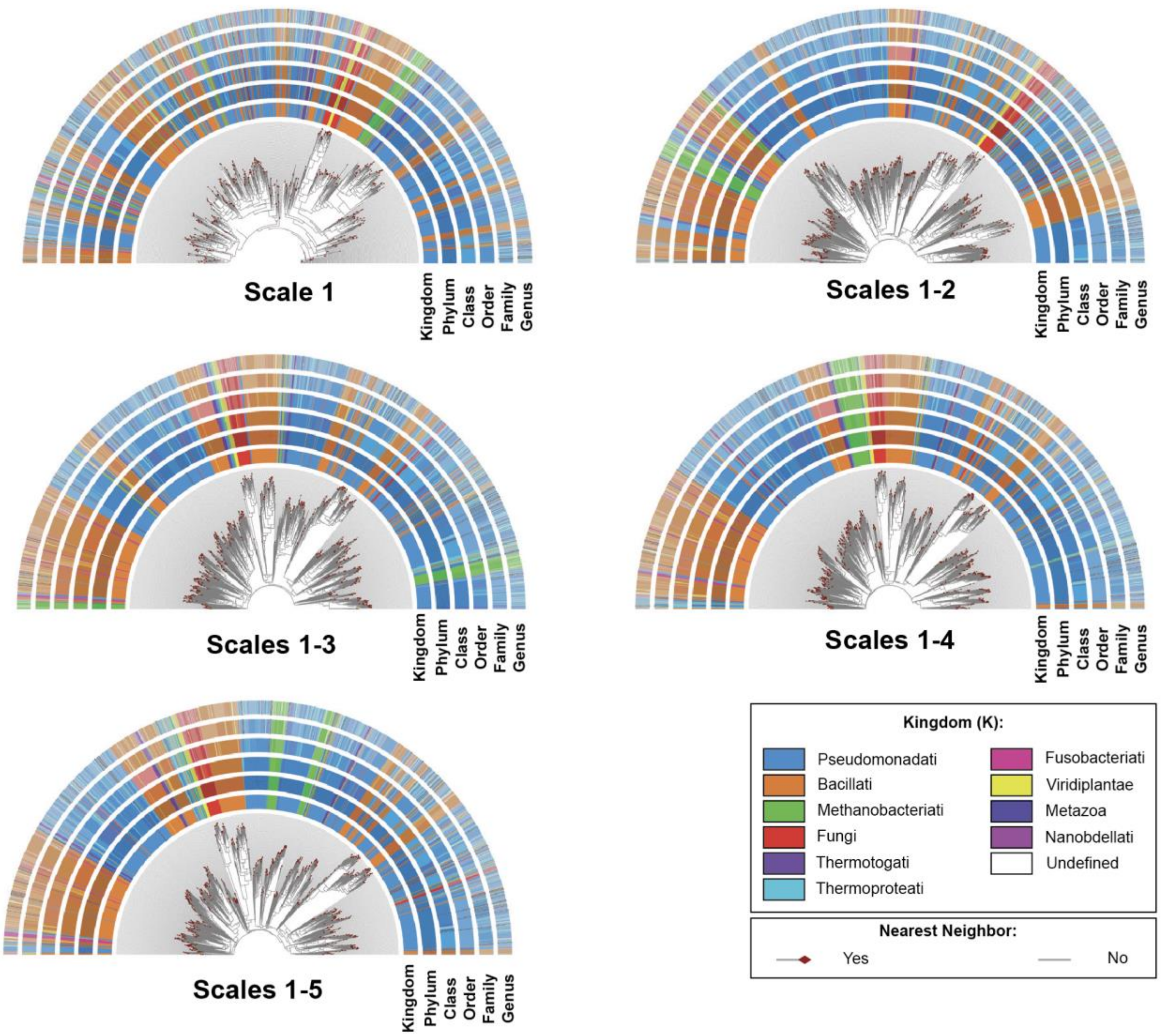

**Figure S8.** Phylogenetic trees of natural CM sequences constructed from considering positions reflected by entropic scale 1 (top left), and up to scale 5 (bottom) using neighbor joining construction. Phylogenetic designation at scale of Kingdom shown in color key and displayed in each tree. Nearest neighbor of CM-LRBM generated synthetic sequences shown in red dots.

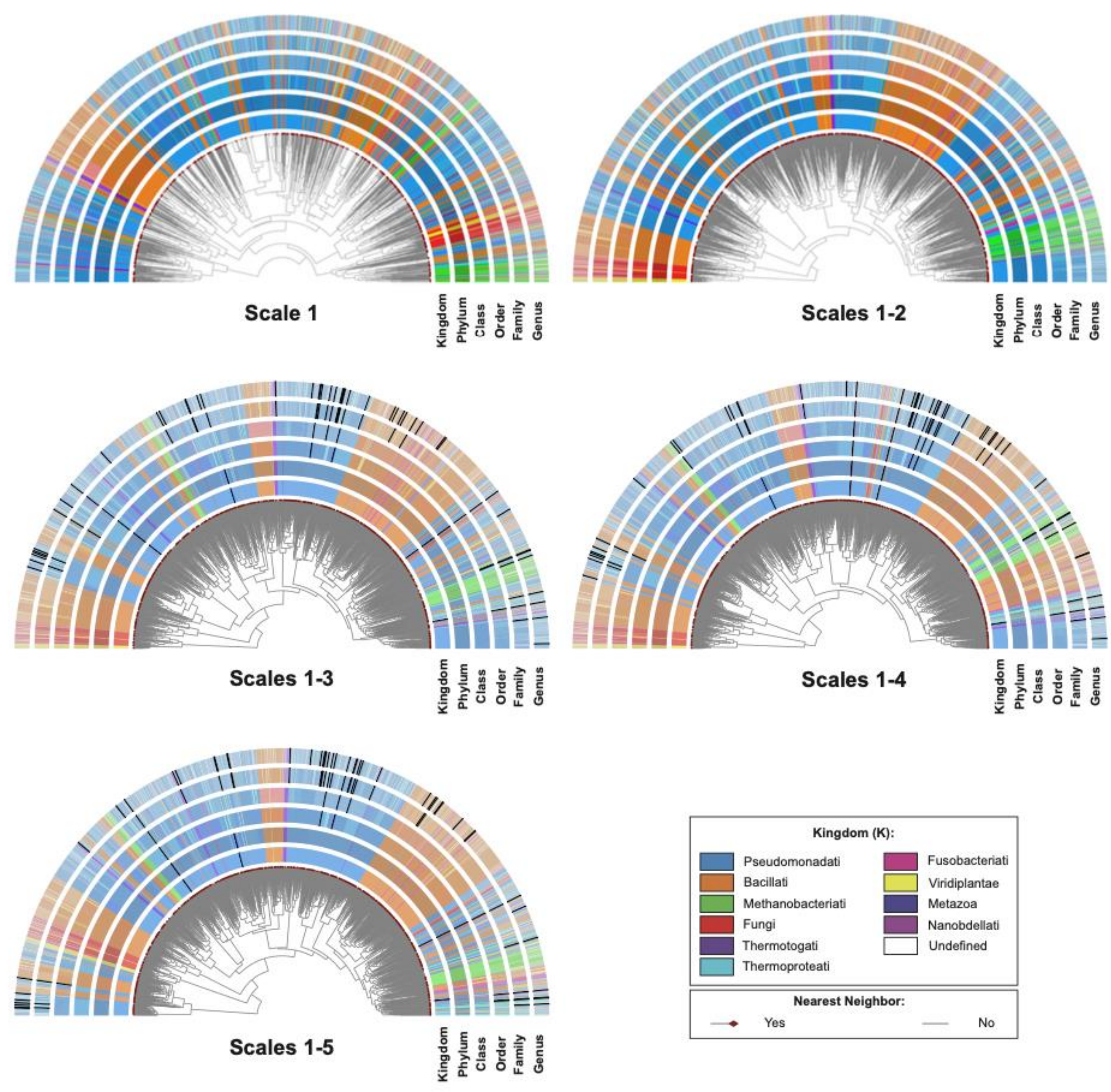

**Figure S9.** UPGMA based phylogenetic trees of natural CM sequences constructed from considering positions reflected by entropic scale 1 (top left), and up to scale 5 (bottom). Phylogenetic designation at scale of Kingdom shown in color key and displayed in each tree. Nearest neighbor of CM-LRBM generated synthetic sequences shown in red dots.

## Methods

### ***Method: Defining scales of entropy***

*Entropy Measurement:*

Given a population, the imprint of evolutionary constraints on any relevant feature describing an individual is left as the conservation pattern observed in that feature's values across the population. If this feature is continuous (say height), the strength of this imprint can be measured through the central tendency and variance across the continuous-valued feature axis. If a feature is under heavy constraint, i.e. locked at a value crucial for survival due to strong selection pressure, the variance in its value observed across the population will be (near) zero. For a feature of relatively low importance for survival, mutational pressure spreads its observed values across the continuum giving rise to large observed variance.

However, for discrete valued features such as genetic and protein sequences, we first need to quantify the notion of central tendency and variance in discrete space for each feature. For this, we borrow the notion of Shannon Entropy from Information theory. Shannon entropy ($S$) is a general measure of disorder or information content of random variable measured from a set of observations. Suppose we have a multiple sequence alignment (MSA) of length '$L$' where each position can take one of $N$ alleles. If $p_j$ is the observed allele frequency for allele '$j$' (of total $N$ alleles) across the k$^{th}$ column of MSA, then the Shannon entropy ($S_k$) of this column is:

$$S_k = -\sum_{j=1}^{N} p_j \log_2 p_j \qquad (2)$$

If the feature is locked at a certain allele value across the entire MSA, only one of the $N$ allele probabilities is 1 while all other $(N-1)$ allele probabilities are 0. In this case, the Shannon entropy is 0. In the other extreme, if all $N$ alleles are observed in equal proportion (each allele probability is 1/N), the Shannon entropy is $\log_2 N$. For a feature with Shannon Entropy $S$, we can now define its effective dimension ($D_{eff}$) - a quantity with a minimum of 1 when a feature is locked at a certain allele (maximally ordered) and a maximum of $N$ when all $N$ alleles are observed in equal proportion across the MSA (maximally disordered).

$$D_{eff} = 2^S \qquad (3)$$

Together, Shannon Entropy and $D_{eff}$ reduce the discrete MSA to a single real number per feature that quantifies the strength of evolutionary constraint, without reference to the specific coupling structure across the system.

In principle, other entropy measures like the Singular value decomposition (SVD) entropy could serve a similar role[50]. SVD entropy is computed from the normalized singular values $\hat{\sigma}_i$of the one-hot encoded MSA column matrix rather than the eigenvalues of its covariance matrix. In practice however, SVD entropy behaves approximately as a Heaviside function in this context, effectively distinguishing only between fully conserved and fully variable features with poor resolution in the intermediate regime. This is because of the SVD spectral compression owing to the square root in the relationship between the singular values and the eigenvalues ( $\hat{\sigma_j} \propto \sqrt{p_j}$ ). Shannon entropy, by contrast, is sensitive to the full spectral structure of the covariance matrix and provides a graded measure that resolves this intermediate regime of evolutionary constraint in addition to the two extremes. This makes Shannon entropy a more natural choice for resolving all regimes of strength of evolutionary constraints.

*Partition of Features into Scales of Entropy:*

After information content of all features is determined using Shannon Entropy, our next goal is to partition of the feature space into scales — collectives of features of similar information level (entropy). By doing so, we convert many raw features into a few effective scales - the feature collectives within which evolution acts through equipotent constraints.

A natural way to perform this partitioning is through the Jenks-Fisher (JF) algorithm, a deterministic algorithm that identifies natural breaks in one-dimensional data[51]. This algorithm comes with a natural objective, the goodness of variance fit (GVF), which measures how well data points cluster around their cluster means relative to the global mean. Given that we are almost always working in the under-sampled regime, rather than naively applying this algorithm to the full MSA, it is better to randomly subsample the MSA, compute the scales and the GVF, and plot the GVF histogram for incremental number of scales. The scale number of diminishing return is determined by identifying where the GVF histograms for two consecutive scale numbers begin to overlap, at which point the lower scale number is selected. Alternatively, the elbow in the plot of mean GVF for incremental scale number can also assign appropriate number of scales.

A more statistically robust alternative is to use a kernel density estimation (KDE) based approach[52]. Here, (300 to 500) random subsampling of 50% of the samples is used to compute the mean and variance of Shannon entropy for each feature in the MSA. Using these means and variances, feature wise normal distributions for entropy are created and combined into a mixture model over the set of all features. The maxima of this mixture model identify the natural cluster centers of scales. Each feature is then assigned to the scale whose center it lies closest to. This approach is more robust than JF in the regime where many features are densely packed within a narrow window of entropy — a situation where JF tends to introduce spurious splits at very fine intervals to inflate the GVF metric. In practice, for protein sequences and sentences, both methods yield near-equivalent partitions with small discrepancies at the scale edges. For image data, however, the high density of features at low effective dimension makes the KDE approach the preferred choice.

For both methods, we impose one additional constraint: single-feature clusters are explicitly merged with their nearest neighbor cluster to prevent information bottleneck. This is because single-feature scales lack the internal interaction parameter structure necessary to mediate statistical influence between the entropic scales on either side. Together, these two methods provide a principled and computationally tractable means of reducing the raw feature space into a small number of effective collectives, each characterized by a representative entropy reflective of the shared strength of evolutionary constraint acting on its members.

The Jenks-Fisher method was built around the 'Jenkspy' python package. The KDE method was built around the python KDE implementation in the signal package's 'find_peaks' function under SciPy library.

***Method: Simulation of an evolved system under hierarchical fitness landscape (Fig. 1B):***

Not all constraints subjected over a population are created equal. Some constraints are extremely crucial for the survival of an individual while others, although not essential for survival of the individual, impart moderate to slight competitive advantage (differential fitness) which might be important for the survival of its lineage. We term the former "strong constraints" and the latter "weak constraints". Because strong constraints make the bulk of differential fitness of an individual, the features (usually gene or residue positions) they govern are often at optimal or near optimal values and homogeneous across the population. This is because, when evolution explores the organism's genotype through genetic variation (mutation and/or sexual reproduction), any variation in such features under strong constraints in an otherwise average individual often leads to a catastrophic loss of fitness ending its lineage. Because of this reason, a proper exploration of this subspace of the features through mutation to find an alternative solution to the constraint is extremely rare thus leading to a dominance of a single or a few solutions across the population. On the other hand, the positions governing a weak constraint are far more flexible because any mutation at these position leads to a small gain or a small loss of fitness. Any organism with a small loss of fitness still holds a significant chance of surviving a probabilistic competition. Thus, several alternative solutions to such a weak constraint are observed across the population. Therefore, across a population, the number of solutions to a constraint can be used as a proxy for the inverse strength of that constraint (relative contribution to fitness).

Along with strength gradient, there can also be a nested hierarchy between these constraints. This hierarchy could be phylogenetic noise — correlation purely of statistical origin because of lineage, or random chance. Or it could be of evolutionary importance — new constraints evolving in the background of ancient constraints as the population gradually adapts to an environment. Unless there is a catastrophic environmental change, these new constraints, compared to the ancient ones, will be weak and satisfying them lends only a small fitness edge to the individual. As an example, all bacteria have cell walls and the variation in this structure is quite minimal across the species. Having a cell wall is thus a strong constraint because a lack of cell wall leads to immediate death of the individual. Having an outer cell membrane and thinner cell walls (gram-negative bacteria) vs a having single thick cell wall (gram-positive bacteria) on the other hand is a less strong constraint. In the background of this constraint, the lineage of gram-negative bacteria developed different modes of motility and complex motor mechanism while the gram-positive bacteria lack such diversity of motility mechanism and are majority non-motile. This diversity of motility mechanism points to it being a weak constraint for survival that evolved in the background of much stronger constraints like having a cell-wall in entire tree of life of bacteria.

To study such a system evolving in hierarchically nested constraint landscape, we built a finite capacity population model stochastically evolving on a fitness landscape with tunable hierarchical epistasis. An individual in this population is described by a genotype of length 16 *over* 4 unique alphabets (alleles). To create a fitness landscape, we partition these 16 features into 4 ordered blocks each governing the effect of a few positions from the genotype on the individual's fitness (see **Fig. 1B**). For simplicity, we assume that any position can only be included in a single block – the explicit effect of that position on fitness is localized to the block including it. We impart evolutionary realism into our fitness landscape by assuming the following:

1. Blocks earlier in ordering code for stronger constraints while latter blocks code for weaker constraints. This is reflected in their fitness contribution.
2. The number of solutions imparting optimal fitness at strong constraint blocks are fewer than the number of options imparting optimal fitness at weak constraint blocks.
3. One way influence among the ordered blocks: blocks earlier in that ordering affect latter blocks but not vice-versa.

For tractability, we model this phenomenon as a binary tree where a block at order '*m*' has $2^{m-1}$ independent constraint sequences imparting optimal fitness. The strength and hierarchy of constraints are programmed through a nested conditional fitness function. If $F_0$, $F_1$, $F_2$, and $F_3$ are the fitness contributions of each block on an individual, then the individual's total fitness is given by a cascading sum of product defined as

$$F_{individual} = F_0 + F_0 \cdot F_1 + F_0 \cdot F_1 \cdot F_2 + F_0 \cdot F_1 \cdot F_2 \cdot F_3 \quad (4)$$

Although this constraint tree can be generated at random, we generate it using a greedy Hamming-maximization procedure such that the constraints are as orthogonal to each other as maximally possible to produce a cleaner separation between fitness effects of each block. The fitness effect ($f_i$) of each position '*i*' within each block is drawn from a normal distribution with an assigned a mean weight and a small noise threshold (10% of the mean weight) to create a variable, noisy fitness contribution across the genome. We set the mean weight for the first two blocks to 1, the third block to 0.5 and the last block to 0. Here, the first block containing s single optimal constraint sequence models an extremely strong constraint. The second block with two optimal constraint sequences but similar weight as first block models a moderately strong constraint. The third block with four optimal constraint sequences models a moderate/weak constraint, and the last block models unconstrained neutral positions.

The fitness contribution ($F_k$) of each block '$k$' on an individual is computed by measuring the hamming similarity between its sequence positions along that block and the ideal sequences at the fitness tree in that block weighted by the fitness effect ($f_i$) of corresponding positions. The total fitness is then computed by the cascading sum of product along all paths of the tree and taking the maximum possible fitness value among all possibilities.

After instantiating out fitness landscape, we initialized a population of 5000 random genotypes and evolve it over discrete epochs using a Moran-type local tournaments[53]. In each epoch of the tournament, we first replaced all individuals with zero fitness with mutated copies of surviving individuals. These zero fitness individuals are the ones with lethal mutations at the first block — the most fitness-critical block. Then we sampled a fraction of the population (set to 30%) and randomly paired them to compete in a fitness based probabilistic pairwise duel. The winning probability was set by a logistic function of the fitness difference between the pairs scaled by a selection-strength (set to 4). The losers of these duels were replaced by mutated copies of the corresponding winners. The mutation was Poisson-type with some average rate of positions being mutated per reproduction event (set to 30%). We ran 10,000 rounds of such tournament at which point the fitness distribution of the population was stable with only minor fluctuation per epoch. The population we got at the end of this process is the final population that stochastically evolved under hierarchical constrained landscape.

## ***Method: Potts Model, LRBM and Profile Models: Energy, Probability, and Sampling***

### *A. One-hot encoding representation:*

Consider a system of length $q$, where each residue position can take one of $m$ possible alphabets. Rather than representing a residue position by the alphabet occupying it, one-hot encoding represents that position as a length-$m$ sub-vector with a 1 at the index corresponding to the occupying alphabet and 0s elsewhere. Concatenating the $q$ sub-vectors, one per residue position, yields a one-hot encoded vector of length $m.q$ for that state (data point). All data states are converted to this representation, since the model operates on it

*B. <u>Potts model, Potts Hamiltonian, energy, and probability of a state:</u>*

The Potts model is an energy-based model with a long history in statistical mechanics[37,54]. Given a system of length $q$, this model considers site-wise propensity of a residue position as well as pairwise couplings between two distinct residue positions through an energy function termed the Hamiltonian. The energy function is then converted to probability using Boltzmann formalism. Thus, for all possible states of the system, the model assigns both an energy value and an associated probability value, forming respectively the energy landscape and the probability landscape. Under standard sign convention, states with lower energy have higher probability - states at the bottom of energy valleys lie at the top of probability hills and are therefore more probable than neighboring states.

Using the one-hot representation introduced above, let a state $S$ be given by sub-vectors $s_1, s_2, \ldots s_q$, one per residue position. A fundamental assumption of Potts model is the symmetry in pairwise couplings - the effect of position *i* on position *j* is the same as the effect of position *j* on position *i*. The Potts Hamiltonian, capturing the site-wise propensity, symmetric pairwise coupling, and the sign convention, is given by:

$$H(s) = -\sum_{i=1}^{q} h_i^T . s_i - \frac{1}{2}\sum_{i \neq j}^{q} s_i^T . J_{ij} . s_j \qquad (5)$$

Here, $h_i$ is a length-m vector of fields at position $i$ that captures the intrinsic preference for each alphabet at that position independent of any other position. Similarly, $J_{ij}$ is a symmetric coupling matrix of dimension $m \times m$ that captures how the alphabet identity at position $i$ influences (and gets influenced by) the alphabet identity at position $j$. The factor of ½ corrects for the fact the coupling structure being symmetric double counts each pairwise interaction - once as $(i, j)$ block and once as $(j, i)$ block. The one-hot encoding of residue positions acts as an indexing mechanism into $h$ and $J$. Because $s_i$ and $s_j$ are one-hot encoded, the terms $h_i^T . s_i$ and $s_i^T . J_{ij} . s_j$ simply select the single field value corresponding to the specific alphabet occupying position $i$ and the single coupling value corresponding to the specific alphabets occupying positions $i$ and $j$. A more compact and operationally useful representation of this Hamiltonian is possible in matrix notation. As noted above, concatenating all sub-vectors $s_i$ for a given state $S$ yields the full one-hot encoded vector $\vec{S}$ of length $m.q$. Similarly, concatenating the field vectors $h_i$ for all positions gives a single bias vector $B$ of length $m.q$. Arranging all coupling matrices $J_{ij}$ gives a single weight matrix $W$ of size $(m.q) \times (m.q)$. This matrix is symmetric by construction with all $(m \times m)$ diagonal blocks representing the self-interaction within a single position set to zero. In this notation, the Potts Hamiltonian compactly becomes:

$$H(S) = -\, B^T . \vec{S} - \frac{1}{2} \vec{S}^{\,T} . W . \vec{S} \qquad (6)$$

Evaluating $H(S)$ for a given state $S$ yields its energy: a single scalar value locating that state on the energy landscape. This energy $H(S)$ for state $S$ is converted into corresponding likelihood $L(S)$ and probability $P(S)$ through the Boltzmann distribution:

$$L(S) = e^{-\,\beta H(S)} \text{ and,}$$

$$P(s) = \frac{1}{\mathcal{Z}} e^{-\beta H(S)} \text{ where, } \mathcal{Z} = \sum_S e^{-\beta H(S)} \qquad (7)$$

Here $\mathcal{Z}$ is the partition function summing over the likelihoods of all possible $m^q$ states. The division of likelihood by this partition function yields the probability $P(S)$ such that the sum of probabilities of all possible states sums to 1. With this form for the probability, lower energy states are exponentially mode probable and higher energy states are exponentially suppressed. Another parameter in the Boltzmann probability distribution is the inverse temperature $\beta$. Inverse temperature controls how sharply probability concentrates around low-energy states. At low temperatures i.e. high $\beta$, the exponential is steep causing the probability mass to be concentrated at the deepest valleys of the energy landscape while at high temperatures i.e. low $\beta$, the exponential flattens with probability spreading more evenly across the landscape. The former restricts higher-energy states while the latter allows higher-energy states to be visited readily. In our work, unless explicitly states, we set $\beta$ to 1.

We next discuss these concepts in the context of LRBM to highlight the similarities and differences between these two formalisms.

### C. *LRBM Model, Hamiltonian, Energy and Probability:*

For the same system of length $q$, the LRBM framework unlike Potts modes doesn't instantiate a generic Hamiltonian form. Instead, its model parameter structures are bespoke to the entropic scales of the dataset being modeled in that they obey the entropic ordering present in the data-features (residue positions). Therefore, scales of entropy are first inferred from the multiple sequence alignment (MSA) of the dataset being modeled. This is done by computing the Shannon entropy at each residue position and applying standard partitioning algorithms such as Jenks-Fisher or kernel density estimation, to the entropic scales – distinct clusters of positions of similar entropy. By construction, each residue position belongs to exactly one entropic scale. Hence, these scales of entropy are disjoint, and their union spans the entire $q$-dimensional state space of the system.

#### 1. *One-hot encoding representation in LRBM Framework:*

Suppose for this system, $k$ scales of entropy are inferred from the dataset. Then, given a protein sequence $S$ of length $q$, computing its energy under LRBM framework begins by projecting $S$ onto each scale of entropy. This is akin to decomposing $S$ into $k$ distinct and disjoint pieces $\{S_1, S_2, \dots S_k\}$, ordered by ascending entropy – $S_1$ contains the $q_1$ lowest-entropy residue positions (scale 1) and $S_k$ contains the $q_k$ highest-entropy residue position (scale $k$). Each piece $S_j$ is then one-hot encoded independently to get $\vec{S_j}$ following the same encoding scheme introduced earlier. This yields a set of $k$ disjoint one-hot encoded sub-sequences that together represent the full sequence $S$.

$$S \rightarrow \{\, S_1, S_2, \dots, S_k\} \quad where, \quad S_j = PROJ_{\text{Scale j}}(S)$$

$$S_i \cap S_j = \phi \quad \forall i \neq j, \qquad and \quad \bigcup_{i=1}^{k} S_i = S$$

$$\vec{S_j} = One - hot\ encode(S_j) \tag{8}$$

2. *LRBM Model Architecture:*

The LRBM model architecture is instantiated with the inferred entropic scales serving as a structural prior on the model parameters. Within this architecture, interactions are only permitted between two immediately adjacent scales of entropy, and only in the direction of increasing entropy — a lower entropy scale influences the immediately higher one but not vice versa.

Suppose scale $j$ consists of $q_j$ residue positions, and the immediately adjacent lower-entropy scale $(j-1)$ contains $q_{j-1}$ residue positions. For any sequence $S$, the $q_j$ positions of $S_j$ interact both amongst themselves and under the influence of the $q_{j-1}$ positions of $S_{j-1}$. Just as in Potts model, we capture these interactions through pairwise couplings and per-position mean propensities. But the adjacent directional connectivity in the order of increasing entropy brings about asymmetry in the coupling matrix.

Concretely, the layer $S_j$ is parametrized by a weight matrix $W_j$ of size $(q_j.m) \times ([q_{j-1} + q_j].m)$ and a bias vector $B_k$ of length $(q_j.m)$. The $(q_j.m)$ rows of $W_j$ correspond to the one-hot encoded positions of $S_j$ being modeled. Its $[q_{j-1} + q_j].m$ columns decompose into two distinct blocks. The first $(q_{(j-1)}.m)$ columns capture the conditional coupling (influence) of the positions of $S_{j-1}$ on $S_j$. The $(q_j.m)$ columns capture the self-coupling among the $q_j$ positions in $S_j$.

3. *LRBM State Energy:*

It is helpful to interpret any element of $W_j$ as encoding a directed relationship: the column index modulo $m$ corresponds to the actor residue position while the row index modulo $m$ corresponds to the residue position being acted upon. With this relationship in place, the two coupling types in $W_j$ can be separated into two distinct matrices: a symmetric square matrix $W_j^{SC}$ for the self-coupling (SC) within scale $j$, and an asymmetric matrix $W_j^{CC}$ for the conditional coupling (CC) from scale $(j-1)$ to scale $j$. As in Potts model, the self-coupling of a position with itself is explicitly zeroed out in $W_j^{SC}$,since a residue position cannot meaningfully interact with itself except through its mean propensity. This mean propensity of each residue position in the absence of any coupling is modeled by the bias vector $\vec{B}_j$. With all these in place, LRBM energy at scale $j$ of this state $S$ is defined as;

$$H_j(S_j) = -\vec{S}_{j-1}^T . W_j^{\mathrm{CC}} . \vec{S}_j \quad - \quad \frac{1}{2} \vec{S}_j^T . W_j^{SC} . \vec{S}_j \quad - \quad \vec{B}_j^T . \vec{S}_j \qquad (9)$$

Here the first term in the sum is the energy contribution due to the conditional coupling (CC) imposed by positions in previous scale $S_{j-1}$ on the current scale's positions $S_j$. The second term terms is the energy contribution due to self-coupling (SC) among positions within the current scale. The third term is the bias energy contribution due to the mean propensity of current scale's positions.

The absence of a factor of ½ in front of the conditional coupling in contrast to the self-coupling term is salient. This is the result of explicitly programmed one-way conditionality between positions in $S_{j-1}$ and $S_j$ with influence going from scale $(j-1)$ to scale $j$ but not in reverse in the LRBM framework. Consequently, unlike the symmetric self-coupling matrix which double counts each position pair within a scale, the conditional coupling matrix visits each cross-scale position pair exactly once. Hence no double-counting correction is needed. We can reformulate the conditional coupling energy as an effective conditional bias ($\vec{B}_j^{CC}$).

$$H_j(S_j) = -\frac{1}{2}\ \vec{S}_j^T.W_j^{SC}.\vec{S}_j - \left[\vec{B}_j + \vec{B}_j^{CC}\right]^T.\vec{S}_j = H_j(\vec{S}_j)^{SC} + H_j(\vec{S}_j|\,\vec{S}_{j-1})^{CC}$$

$$\text{where, } \vec{B}_j^{CC} = \vec{S}_{j-1}^{\,T}.W_j^{CC} \ ,$$

$$H_j(S_j)^{SC} = -\tfrac{1}{2}\ \vec{S}_j^T.W_j^{SC}.\vec{S}_j - \vec{B}_j^T.\vec{S}_j \ ,$$

$$H_j(S_j|\,S_{j-1})^{CC} = -\ \vec{B}_j^{CC\,T}.\vec{S}_j \qquad (10)$$

This can be thought of as the scale $(j-1)$, based on the content of $\vec{S}_{j-1}$ and $\vec{S}_j$, applying an effective conditional bias $\vec{B}_j^{CC}$ to tilt the self-coupling energy landscape $H_j{}^{SC}$. This $j^{th}$ energy landscape thus becomes dynamic due to this state dependency. In contrast to Potts model where any state $S$ is only represented by a single scaler $H(S)$ in one global static energy landscape, a LRBM with $k$ scales represent the same state with $k$ scaler valued energies, $\{H_1(S), H_2(S), ..H_k(S)\}$, in $k$ distinct state-dependent dynamic energy landscapes that are chained across the entropic scales through the conditional coupling. Only the lowest entropy scale has a tilt free and static energy landscape $H_1$ because there is no scale of lower entropy to influence it. In LRBM, the sum of energies of all $k$ entropic scales is can be thought of as the total energy for any given state S. However, unlike Potts model, operationally it is these $k$ scaler valued energies, rather than the total energy, that are used for state evolution.

For completeness, we also present the microscopic form of LRBM Hamiltonian ($H_j$) for j[th] energy landscape. Here the first term represents the per-site propensity in the scale $j$, the second term (with ½) represents the symmetric self-coupling between positions in scale $j$, and the last term represents the one-way conditional coupling from scale $(j-1)$ to positions in scale $j$.

$$H_j(S) = -\sum_{m \in S_j}^{q} h_m^T . s_m - \frac{1}{2} \sum_{\substack{m \neq n \\ m,n \in S_j}}^{q} s_m^T . J_{mn} . s_n - \sum_{\substack{m \in S_j \\ N \in S_{j-1}}}^{q} s_N^T . J_{Nm} . s_m \qquad (11)$$

4. *LRBM State Probability:*

Given the j[th] scale LRBM energy $H_j(S)$ for state $S$, the corresponding likelihood $L(\vec{S}_j)$ and probability $P(\vec{S}_j)$ are given by the Boltzmann distribution over $S_j$ conditioned on $S_{j-1}$:

$$L(S_j |\, S_{j-1}) = e^{-\beta\, H_j(S)} = e^{\left(-\beta\, H_j(\vec{S}_j)^{SC}\right)} . e^{\left(-\beta\, H_j(\vec{S}_j |\, \vec{S}_{j-1})^{CS}\right)}$$

$$\mathrm{P}(S_j |\, S_{j-1}) = \frac{1}{\mathcal{Z}_j} . e^{-\beta H_j(S)} = \frac{1}{\mathcal{Z}_j} e^{\left(-\beta H_j(\vec{S}_j)^{SC}\right)} . e^{\left(-\beta H_j(\vec{S}_j |\, \vec{S}_{j-1})^{CS}\right)} \qquad (12)$$

Here too, we set the inverse temperature parameter $\beta$ to 1. Here, $\mathcal{Z}_j$ is the partition function computed over all $q_j.m$ possibilities in j[th] entropic scale. It doesn't consider the possibilities at $(j-1)^{\text{th}}$ entropic scale because by zeroing out the reverse connection between the two scales, we are explicitly enforcing a direction of entropic causality— all residue positions $(j-1)^{\text{th}}$ entropic scale have already crystallized and thus form a background in which the residue positions of $j^{\text{th}}$ entropic scale fluctuate. There is no probability of $S_j$ in isolation, only $P(S_j|S_{j-1})$. Each scale's probability is only defined relative to the scale immediately lower in entropy. Considering this entropic causality in LRBM framework, for any state $S$ we serially evaluate the probability along the entropic scales beginning with the first entropic scale ($S_1$). Thus, the total probability (across all scales of entropy) for the state $S$ in the LRBM framework is:

$$P(S) = P(S_k|S_{k-1}) \dots P\left(S_j|S_{j-1}\right)..P(S_2|S_1).P(S_1) \;=\; P(S_1).\prod_{j=2}^{k} P\left(S_j|S_{j-1}\right) \qquad (13)$$

We note that the total probability $P(S)$ in LRBM architecture has a nested structure: the probability of the full sequence is built from the probability of $S_1$ on its own, times the probability of $S_2$ nested within the context of $S_1$, times the probability of $S_3$ nested within the context of $\{S_1, S_2\}$ ,and so on all the way to the highest entropy scale $k$.

*D. <u>Sampling in Potts and LRBM Framework:</u>*

In the Potts model, sampling a state from the learned distribution proceeds by initializing a completely random state across all $q$ positions, computing its global energy $H(S)$. A probabilistic Markov Chain Monte Carlo (MCMC) walk is then performed in which ideally a single or practically a small subset of residue positions is proposed to change at each step based on the energy differences these changes bring about. Suppose $S$ and $S'$ represent the state before and after these changes. The acceptance probability of any change to a state is equal to the ratio of probabilities $P(S')$ to $P(S)$. These probabilities however are computationally intractable to evaluate due to astronomically large number of possible states needed to be accounted for to calculate the partition function $\mathcal{Z}$. MCMC sidesteps this by using the ratios of likelihoods $L(S')$ and $L(S)$ instead which yield the same result as the ratio of probabilities without requiring $\mathcal{Z}$. Proposed changes that lower the energy (raise the likelihood) are preferentially accepted to those that raise the energy (lower the likelihood) in accordance with this ratio. Repeating this procedure over many steps, the initial state slowly descends in the single, static Potts energy landscape until it settles (relaxes) into a valley.

Sampling in the LRBM framework departs from Potts in a fundamental way because of the scale-wise conditionally tilted structures of its energy landscapes and the resulting nested probability distribution described above. Rather than relaxing the entire sequence at once, LRBM sampling proceeds serially*,* scale by scale, in the direction of increasing entropy respecting the same entropic causality used in computing the total probability $P(S)$.

To start sampling procedure from a completely random initial state $S$, we first start with $S_1$— the residue positions pertaining to the 1st entropic scale. Since there is no layer influencing this scale, sampling here is exactly equivalent to Potts model $H_1(S)$ with sampling described above. Once this is complete, sampling for the 2nd entropic scale begins by masking $S_2$— the residue positions belonging to the 2nd entropic scale while the residue positions of $S_1$, the immediately preceding lower entropy scale, are held frozen at their current values. With $S_1$ fixed, the 2nd scale's energy $H_2(S)$ is computed, and an MCMC walk is performed only over the energy landscape associated with 2nd entropic scale using the same likelihood-ratio acceptance rule described above but now restricted to this energy landscape. During each step of this walk a single or a small subset of residue positions are changed until a low energy (high likelihood) configuration for $S_2$ is found conditioned on the fixed background $S_1$. In this way, for each trajectory, the peaks and valleys of $H_2$ shift depending on the specific configuration of $S_1$ inherited from the previous scale. This is the central difference between LRBM and Potts model. Rather than one global, state-independent landscape shared by all positions, LRBM sampling relaxes the state at each scale onto its own landscape, exclusively tilted by whatever configuration was fixed at the preceding scale.

This procedure is repeated serially across all scales, beginning at the tilt-free lowest entropy scale $S_1$ all the way up to the highest entropic scale $S_k$. At the end of each scale $j$, the newly relaxed $S_j$ becomes the fixed context for relaxing the next scale $S_{j+1}$. This serial sweep under gradually accrued context is the sampling-side realization of the LRBM's nested probability structure noted above.

*E. Training the LRBM: Layer Restricted Contrastive Divergence:*

Having established how a state is sampled serially over the scales under the LRBM framework, we now turn to how the model parameters themselves, namely $W_j^{SC}$, $W_j^{CC}$, and $B_j$ at each scale are learned from data (the provided MSA). First, we instantiate the LRBM with the appropriate structures (from entropy scales) enforced on the model parameters. The actual values of the parameters are set to random except those that are explicitly zero by LRBM assumptions.

We implement a layer-restricted training scheme in which the parameters governing each scale $j$ are trained as their own Boltzmann-type (Potts) model tilted by the effective conditional bias $B_j^{CC}$ arising from the influence of previous scale. For each scale $j$, all parameters are tuned by contrasting the positive (Hebbian) samples against negative (anti-Hebbian) samples generated for every data point in the MSA. For a single data point $S$, the positive sample at scale $j$ is its projections onto scales $j$ and $(j-1)$ i.e. $\{S_{j-1}, S_j\}$. The corresponding negative sample is obtained by randomizing scale $j$ positions of S and sampling from the current state of the model while under the influence of $S_{j-1}$ until the state relaxes to get to $\{S_{j-1}, S_j^{'}\}$ — a valley of the current energy landscape.

For a well-trained model in principle, this relaxed negative sample $S_j^{'}$ should match the positive sample $S_j$ for that data point, and any metric used to contrast these two samples should vanish. However, because the state evolution in the energy landscape is governed by a probabilistic rule (likelihood-acceptance rule) rather than a deterministic one, we do not compare samples on a per-datapoint basis. Instead, we compare the ensembles of positive samples to the ensemble of negative samples generated from the training data (or its random subset in practice). Under the maximum likelihood framework, the appropriate metrics for the ensemble-wide contrast are the single-site frequencies of residues at each position within scale $j$ for mean-propensities and pairwise frequencies of residue pairs between positions within scale $j$ for self-coupling as well as across scales $j$ and $(j-1)$ for conditional coupling. Concretely, these frequencies define the parameter gradients used for parameter update directly, without requiring explicit computation of the partition function. These gradients are:

$$\Delta B_j \;\propto\; \langle \vec{S}_j \rangle_{\text{pos}} - \langle \vec{S}_j \rangle_{\text{neg}}$$

$$\Delta W_j^{SC} \;\propto\; \langle \vec{S}_j \,.\, \vec{S}_j^{\,T} \rangle_{\text{pos}} - \langle \vec{S}_j . \vec{S}_j^{\,T} \rangle_{\text{neg}}$$

$$\Delta W_j^{CC} \;\propto\; \langle \vec{S}_{j-1} . \vec{S}_j^{\,T} \rangle_{\text{pos}} - \langle \vec{S}_{j-1} . \vec{S}_j^{\,T} \rangle_{\text{neg}} \qquad (14)$$

Here, $\langle \cdot \rangle_{\text{pos}}$ denotes an average taken over the positive (Hebbian) ensemble — the empirical single-site and pairwise frequencies observed in the training MSA. Similarly, $\langle \cdot \rangle_{\text{neg}}$ denotes the corresponding average over the negative (anti-Hebbian) ensemble, obtained by relaxing the randomized scale-$j$ positions under the model's current parameters. These updates follow directly from gradient ascent on the log-likelihood of the data under the model (a derivation we do not repeat here)[17]. Intuitively, these updates can be thought of as raising biases and weight parameters for residues and pairs under-represented in the model relative to the data and lowering for those that are over-represented. This nudges the model's negative ensemble toward the statistics of the positive ensemble at each scale, one contrastive divergence step at a time.

To improve mixing across this conditional energy landscape, we exploit the inverse temperature-dependence of $P(S_j \mid S_{j-1})$ introduced above via parallel tempering. Several Gibbs chains each loaded with a random initial state restricted to the scale-$j$ positions, are run in parallel across a ladder of inverse temperatures, with linear spacing between rungs performing best in our experience. After all chains advance by a fixed number of Gibbs iterations, swaps are attempted between adjacent-temperature chains using the standard replica-exchange acceptance rule, based on the energy difference between the two chains and the difference in their inverse temperatures[55]. This allows the slow-mixing valley-finding low-temperature (high inverse temperature) chains sample low energy states faithfully by occasionally inheriting states explored by high-temperature (low inverse temperature) chains which mix quickly and sample more broadly across the landscape. The negative ensembles thus obtained better reflect the true model distribution at each scale, which in turn produces more accurate and robust gradient estimates during training.

One the positive and negative ensembles are formed, the model parameters gradients $\Delta B_j$, $\Delta W_j^{SC}$, and $\Delta W_j^{CC}$ are computed from the differences between the ensemble averaged frequencies. These gradients are scaled by a learning rate $\eta$ and applied additively to the current parameter at each training iteration.

$$B_j \leftarrow B_j + \eta\,\Delta B_j, \qquad W_j^{SC} \leftarrow W_j^{SC} + \eta\,\Delta W_j^{SC}, \qquad W_j^{CC} \leftarrow W_j^{CC} + \eta\,\Delta W_j^{CC} \qquad (15)$$

If the learning rate is too large, the parameters often oscillate around rather than settle into their optimal values, while too small a value makes training excessively slow. This process of sampling, contrasting ensembles, and updating parameters is repeated iteratively until a stopping criterion is met. This criterion is typically when the contrastive metrics (single and pairwise-site frequency differences between the positive and negative ensembles) fall below a pre-specified tolerance, or when these differences plateau across successive iterations without further improvement, indicating that the model's sampled distribution has converged to the training data's distribution at each LRBM scale.

*F. Profile Model: Energy, Probability and Training*

A profile model is the simplest special case of the Potts model obtained by setting all pairwise couplings to zero ( $J_{ij} = 0$ for all $i \neq j$)[56]. What remains is a model that captures only the site-wise, independent propensity of each residue position without any coupling between positions. Thus, each position is treated as statistically independent of from others. Using the same one-hot representation and notation established for the Potts model, the Hamiltonian of a profile model for a state $S$ reduces to just the field (bias) term and the corresponding probability follows the same Boltzmann formalism:

$$H(s) = -\sum_{i=1}^{q} h_i^T . s_i$$

$$P(s) = \frac{1}{\mathcal{Z}} e^{-\beta H(S)} = \frac{1}{\mathcal{Z}} \prod_{i=1}^{q} e^{h_i^T . s_i} \qquad (16)$$

Without a coupling term, the exponential factorizes cleanly across positions, and correspondingly the partition function ( $\mathcal{Z}$) factorizes as well into a produce of individual position partition functions ( $\mathcal{Z}_i$).

$$\mathcal{Z} = \prod_{i=1}^{q} \mathcal{Z}_i, \text{ where } \mathcal{Z}_i = \sum_{\alpha=1}^{m} e^{h_i(\alpha)} \qquad (17)$$

Here $\alpha$ in $\mathcal{Z}_i$ sums over the $m$ possible alphabets at position $i$ alone. This means that the probability of a state under a profile model is simply the product of $q$ independent, per-position probabilities. Since each position is independent, a profile model is only required to reproduce the single-site frequencies of residues observed at every position in the training data. Because of this modularity, although probabilistic, profile model has exact closed form solutions for the model parameters. If $\langle s_i(\alpha) \rangle_{\text{data}}$ denotes the empirical frequency of alphabet $\alpha$ at position '$i$' averaged over the training MSA, setting the model's per position probability equal to this empirical frequency gives the exact value for the model's field parameter $h_i(\alpha)$ that governs it.

$$\frac{e^{h_i(\alpha)}}{\mathcal{Z}_i} = \langle s_i(\alpha)\rangle_{\text{data}} \quad \Rightarrow \quad h_i(\alpha) = log\langle s_i(\alpha)\rangle_{\text{data}} + c_i \qquad (18)$$

The per position constant ' $c_i$' which comes about because of $log(Z_i)$ is inconsequential for model's generative capacity since the partition functions cancel out while sampling using likelihood-based ratios. These inconsequential constants of the model are collective termed 'gauge freedom'. They are typically fixed by the user's choice of gauge. Some typical examples include the 'zero reference' gauge —setting $h_i(\alpha) = 0$ for a reference alphabet $\alpha$, and the 'zero sum' gauge — setting $\sum_\alpha h_i\,(\alpha) = 0$ .

***<u>Method: Recursive Tokenization, End Padding, and Color Compression</u>***

Unlike other datasets in our paper, sentences provided a unique two-fold challenge. On one hand, although sentences form discrete datasets, the number of possible words in English language are in the order of 600,000. If we model each position in the sentence with 600,000 internal degrees of freedom, creating a modest sentence of length 10 required handling a vector with 6,000,000 degrees of freedom with a model of size in the order of squared of this value. On the other hand, given a modestly sized training dataset, most of these words will never be observed. And for those that are observed, the number of observations will be small. Only a small fraction of commonly used words will show high multiplicity across the dataset. Hence, even if we create an all-encompassing model, most of the parameters will be poorly tuned because of this fact. To surpass these challenges, we make the following four design choices:

*A. <u>Data Aware Model:</u>*

Rather than modeling all 600,000 unique words in the English language, we constrained the LRBM to model only the uniquely observed words in the animal Wikipedia training corpus sentences. This drastically dropped the total number of unique words to be modeled to 2855 words.

*B. <u>Recursive Tokenization:</u>*

Language inherently is flexible in that same root word can form numerous unique words depending on the prefix or suffix added to it. To exploit this property, we implemented a recursive, morphology aware tokenization scheme that builds a sub-word vocabulary directly from the corpus of unique words observed in the training set. The tokens (collection of sub-words and unsplit words) remaining at the end form the actual degrees of freedom for sentence at any position rather than the unique word corpus. To find these tokens, we first built a vocabulary dictionary recording each unique word's frequency in the dataset. We alphabetically grouped this unique word set and sorted by length within each group to speed up later root word lookups. We also implemented user-supplied word protection scheme by making sure that any word appended to this protected list is not to be split further and remain as a single token.

For each word in this list, we attempt three competing decomposition strategies to discover the construction morphology and adopt whichever decomposition scores highest in our heuristic scoring scheme. The three strategies are: a double-word split, a prefix-word split, and a word-suffix split. A double-word split searches all internal cut points for a pair of sub-words that are both unique words in the vocabulary list. A suffix split checks across all splits of the word to find a decomposition with a root word in the vocabulary list plus a suffix. Priority is given to splits with longer root words through scoring when multiple splits are possible. A minimum protected root word length criterion (initially set to 3) as well as ratio criteria of suffix length being less than or equal to 0.6 of root length is set to prevent splits where suffix length exceeds the root length. A similar implementation is done for prefix splits of the word to find a decomposition with a prefix and a root word in the vocabulary. For both prefix and suffix splits, we also enforced a forbidden-affix list to avoid nonsensical splits. These lists were built after evaluating a first splitting attempt without any forbidden list.

For any word in the unique vocabulary list, we attempted all three strategies unless it was a protected word. Highest priority was given to the double-word split by constructing a score function that rewarded for the length of both sub-words. Prefix and suffix splits were rewarded only for the length of the root word. The split with the highest scores were added back to the unique vocabulary list. We then recursed on the resulting sub-word(s) using the same procedure, incrementing a conservativeness counter that raises the minimum protected root word length criterion by 1 at each recursive step. This was done to avoid extremely finer splits with each recursing. This recursion was performed until no further valid split is found, yielding a full decomposition of every corpus word into a sequence of roots, prefixes (marked with a trailing hyphen), and suffixes (marked with a leading hyphen). We stored this decomposition in a token dictionary and flattened it to obtain the final token vocabulary along with the sets of unique prefixes and suffixes.

For every affix (now distinguished by hyphens), we next examined the number of times we observed its usage in the dataset by measuring its frequency as well as tabulating its context (distinct roots). Affixes that occurred in fewer than a minimum number of distinct contexts or below a minimum total frequency were flagged as low quality, added to the forbidden-affix lists. A second fresh pass through the recursive tokenization was performed with this list enforced on top of all previous criteria to get the final token dictionary that housed any unique word as well as its split morphology based on tokens. This second pass prevented idiosyncratic, corpus-specific fragmentation from being retained in the token list.

Although extremely heuristic driven, this recursive tokenization procedure further compressed the unique vocabulary size by roughly 25% (2855 to 2156). The scoring function heuristics and decision trees for the splits were constructed over several iteration of trial and error and we envision the possibility of several improvements in this pipeline.

C. *Length selection and End-Padding:*

We next transformed the training data from word space to token space using the token dictionary. We then selected a length cutoff with a maximum of 17 tokens in any sentence. This retained 91% of the training corpus sentences with bulk of the sentences having a token length of 10. We next needed a way to align these sentences and handle the variability in their length to create a sentence-MSA. Rather than a complicated grammar-based alignment, we opted for a simpler front alignment for all sentences. This led to the length variability showing up at the end of the sentences which we handled by end padding with a dummy hyphen symbol.

D. *Color Compression:*

Rather than modeling all 2156 unique tokens at any given word position, we limited to explicitly modeling only the unique tokens observed at that position along with a single dummy token where all the unobserved tokens are mapped. This is known as color-compression in Potts model literature[57]. This dummy token allows the model to evaluate out-of-sample sentences which might contain a few previously unseen tokens. For sentences, color compression is not only pragmatic but also crucial given the fact that we have an uneven entropy distribution (see **Fig. 1C**). Positions at the beginning and end of sentences are highly constrained with only a few unique tokens as opposed to positions at the center of the sentences where the number of unique tokens observed are several times larger. Color compression thus allows for a pragmatic distribution of parameter resources by allocating a lot of model parameters to model the variable portion and constraining a few parameters to model the conserved portion of the sentences in the training corpus.

***Method: AlphaFold- Rosetta- USAlign Pipeline:***

Using AlphaFold 3 we first folded all 1130 natural proteins (training data) and all five levels in the 150 LRBM trajectories[11]. The resulting structures were saved for downstream Rosetta analysis. The predicted Template Modeling (pTM) scores and the Interface predicted Modeling (ipTM) scores were noted.

The AlphaFold structures were found to hold heavy steric clashes (Lennard-Jones repulsion) according to Rosetta's physics-based model for both natural as well as the LRBM trajectories. To relieve the structures of this conformational stress (see **Fig. S6A**), a relaxation algorithm was run in Rosetta[58]. In this algorithm, firstly, a virtual root residue fixed at the origin was added to the structure to anchor its absolute position as well as orientation during the minimization thus preventing whole-body drift as well as rotations. Following this the structure was relaxed up to 200 iterations using full Cartesian space 'FastRelax' with 'REF2015_CART' all-atom energy function[59,60]. A harmonic coordinate constraint with weight of 2 and deviation of 0.5 Å was also applied to every atom. This constraint imposes a continuous quadratic penalty with the given weight when atoms go beyond this deviation thus permitting a smooth, local relief of structural strain and steric clashes while discouraging a sudden or violent conformational rearrangement.

The total energy being minimized during Rosetta relaxation is a sum of all-atom energy function and the energy due to this harmonic constraint. The weight and deviations for the constraint were chosen after testing over a combination of values on five natural protein's structures. Smaller values of weights and larger values of deviations yielded structures with very similar properties but with a longer run-time (iterations). This relaxation procedure was repeated three times independently per structure. The post-relaxation quantities (Binding Free Energy, Steric Clash, Total Rosetta Energy) reported represent the mean over these three relaxed structures.

We next quantified the structural divergence of each relaxed structure against the relaxed *E. coli* Chorismate Mutase reference structure using per-residue Cα root mean squared deviation (RMSD) in US-align command line executable program[61]. Each structure was first structurally superposed onto the E. Coli reference using US-align in multi-chain mode (-mm 1) to correctly handle the homodimer complex. Following this, per-residue Cα RMSD was computed by matching residues between the aligned mobile and reference structure by residue number. To resolve the chain-level ambiguity, we first split the per-residue RMSD array into two chain halves, selected the half with lower mean RMSD, and mirrored it symmetrically to the remaining chain. This was necessary for a conservative approximation of RMSD especially for sequences earlier in the LRBM trajectory where divergence is high. The RMSD for residues where no corresponding Cα was found in the compared structure due to length mismatch was taken as 'NaN' ('Not a number').

We repeated this procedure across all structures in the ensemble and tabulated the RMSD at each residue position for all 192 residues of the homodimer. This yielded a matrix of shape 1130 X 192 for the naturals as well as 150 X 192 for each level in the trajectory. We next computed column-wise NaN-skipped mean and NaN-skipped standard deviation across these matrices to get the mean and standard deviation of Cα RMSD at each residue position. For each ensemble, this resulting per-residue mean RMSD profile was mapped onto the B-factor channel of the E. coli reference structure file for visualization of the structural divergence map across of an ensemble relative to the reference structure using UCSF ChimeraX[62]. For Chorismate mutase, all structural graphics, solvent accessible surface area calculation, and molecular rendering figures were also generated using UCSF ChimeraX.

***<u>Method: PCA Tree construction and sequence placement in the tree</u>***

To place the generated LRBM sequence (or a previously unseen sequence) in the space of reference natural sequence dataset in a way that respects the scales of entropy, we built a PCA based pipeline for tree construction and data projection onto the tree. What does it mean for a constructed tree diagram to respect the scales of entropy? Firstly, because LRBM scales are nested from low to high entropy, any high entropy scale can only exist in the background of all preceding low entropy scales. Hence, to construct a tree at any entropic scale, we must consider all sequence positions that come under said entropic scale and all the preceding scales. Secondly, to account for the ordering of entropy when computing distances for the tree, we weigh the PCA projected distances onto each principal component by the associated explained variance before summing the contribution of that component to find agglomerative distances. Since low entropy scales dominate PCs with high explained variance (major PCs), they are heavily weighted and thus become the major contributors of agglomerative distance. However, since low entropy scales stay near static across the dataset, they also add near-equivalent contribution across the dataset unless a datapoint happens to be a deviant at these scales. This heavily accentuates any deviation in a datapoint at low entropy scales. High entropy scales on the other hand represent idiosyncrasies in the dataset and are encoded in the minor PCs which have low explained variances. When weighted by the explained variances, the contribution of such PCs to the agglomerative distances is small but different across the dataset. Philosophically, this means that any deviation at low entropy scales is a genuine signal for a unique solution to the problem the ensemble evolved to solve. But upon lack of variance in low entropy scales of a dataset, the high entropy scales become the drivers of solution to evolutionary problem.

To construct the PCA tree at any entropic scale with consideration, we first projected the compact reference data (non-one-hot encoded) to the scale and all preceding scales of entropy. This column-wise subset of the dataset was then one-hot-encoded to an alphabet set of this projected dataset or a larger user-supplied superset of alphabets. If the user-supplied alphabet set was smaller than the projected dataset's alphabets, a union of the two sets was taken as the alphabet set. Although the reference MSA is unique, the column-wise subset obtained for especially lower entropy scales can contain a lot of repeated sequences if most of the variation lie at higher entropy scales. Therefore, to avoid duplicate sequences dominating the principal components, we first purified the reference data to only unique observation and fitted PCA on this unique-only subset of one-hot-encoded reference data.

The number of principal components ‘n’ was capped at the smaller of the sample size and feature count used for fitting. After fitting, we computed the projection of each reference data (regardless of its uniqueness status) from the column-wise subset onto each PC. This projection is the unweighted coordinate of that datapoint along that PC. After doing this operation for a given datapoint across all ‘n’ principal component, we got a n-valued coordinate that uniquely places the datapoint in the ‘n’-dimensional PC space. We next computed a pairwise weighted-Euclidean distance matrix in this n-dimensional PC space over all projected reference data pairs using a standardized Euclidean metric with per-component weight given by the associated explained variance of that component. The distance matrix thus obtained from the reference data was used to compute a reference tree using either standard neighbor-joining algorithm or UPGMA construction. We also computed a global centroid of the reference tree by taking a mean of the unweighted PC coordinates of the reference dataset along each PC axis. This global centroid was later used to place any new sequence onto the reference tree.

To place a new sequence onto this fixed reference tree, we first projected it to the scale and all preceding scales of entropy and one-hot encoded the projection using same alphabet set and ordering as the reference data. We next found its nearest-neighbor reference leaf under the same pairwise weighted distance metric as above. For placement on a reference UPGMA tree, we simply attached the new sequence as a sister leaf of the nearest neighbor with equal branch length directly from that neighbor’s parent. This produces a consistent with ultrametric structure of UPGMA trees. For neighbor-joining reference trees, we computed the weighted Euclidean distance of this new sequence as well as the nearest reference leaf to the global centroid. These three distances- new to nearest neighbor, nearest-neighbor to centroid, and new to global centroid form a triangle. We next use law of cosines to solve for the angle at the nearest-neighbor vertex of this triangle. This angle is then used to decompose the new to nearest-neighbor distance into a perpendicular and a parallel distance component with reference to the nearest neighbor branch using standard trigonometry. This parallel component is used as the position to place an intermediate node where the nearest neighbor and the new sequence splits. The perpendicular component is used as the new sequence leaf's branch length. Although more involved than UPGMA option, the geometric option we described above is a more globally consistent option when editing a neighbor-joining reference tree. We implemented both options in our pipeline to handle both neighbor-joining trees and UPGMA trees. The resulting reference or augmented trees was serialized to Newick format for downstream use or visualization.

The PCA calculation was performed using 'scikit-learn' pca package and the weighted Euclidean distance matrix calculation was performed using the weighted pairwise distance implementation (spatial.distance.pdist) implementation under 'scipy' package by supplying the explained variance as weights along each PC direction. The reference tree construction was performed using Bio Python. The tree visualization and colorization were performed using 'pyCirclize' python package.

***<u>Method: Chorismate Mutase Assay</u>***

A. <u>Gene Construction</u>

340-mer oligonucleotides were designed, each containing a unique coding sequence for an artificial AroQ Chorismate Mutase (CM) protein optimized for *E. coli* codon usage and flanked by primer-annealing sequences and NdeI and XhoI restriction sites. This pool of oligonucleotides was amplified by PCR using KAPA HiFi DNA Polymerase and primers GSP_2 (5'-AAACCGGAGCCATACAGTAC-3') and GSP_107 (5'- CCGTGCGACAAGATTTCAAG-3'), followed by gel purification of the expected product using the QIAquick Gel Extraction Kit. Amplified genes and the pKTCTET-0 backbone were digested with NdeI and XhoI, gel purified and ligated overnight at 16°C at a 1:3 backbone-to-insert molar ratio using T4 DNA ligase (New England Biolabs). The resulting plasmid library was transformed into electrocompetent *E. coli*

DH10β cells (New England Biolabs) at a library coverage of >1000 cells per sequence. The pooled transformants were expanded overnight in LB medium containing 100 μg/mL ampicillin and harvested for plasmid purification. The resulting plasmid library was then diluted to 0.1 ng/μL to minimize the risk of multiple transformation and transformed into CM-deficient KA12 *E. coli* cells harboring the auxiliary pKIMP-UAUC plasmid, again yielding >1000-fold coverage per sequence. Transformed cells were expanded overnight in 100 mL LB medium containing 100 μg/ml ampicillin and 30 μg/ml chloramphenicol, and aliquots were prepared by mixing equal volumes of culture and 33% glycerol, and stored at -80°C.

B. Chorismate Mutase Selection Assay

To quantify the relative activity of the CM libraries used in this study, we adopted a previously described Chorismate mutase selection assay[46]. First, frozen glycerol stocks of KA12/pKIMP-UAUC cells carrying pKTCTET-variant libraries (designed, natural, standard curve) were revived separately by shaking overnight at 30°C in LB medium supplemented with 100 μg/mL ampicillin (amp) and 30 μg/mL chloramphenicol (cam). Each culture was then diluted to an $OD_{600}$ of 0.045 in non-selective M9c medium containing the same two antibiotics plus 20 μg/ml each of L-phenylalanine and L-tyrosine (M9cFY) and grown at 30°C to an $OD_{600}$ of ~0.2. The three cultures were then combined to a total volume of 50 mL at relative representation ratios of 1:1:4 for the designed, natural, and standard curve libraries respectively. This combined culture was subsequently washed with minimal M9c medium (lacking phenylalanine and tyrosine), and a fraction was taken to begin the selective growth. To generate the pre-selection 'input' sample for later analysis, the remaining cells were inoculated into 2 mL LB medium containing amp and cam, followed by overnight growth at 37°C and plasmid purification. The fraction taken for selection was diluted to an $OD_{600}$ of 0.0001 in 1 L M9c medium containing amp and cam, as well as 3 ng/mL doxycycline to induce library expression under the $P_{tet}$ promoter. This culture was grown at 30°C for 24 hours and resulted in a final $OD_{600}$ of <0.1. Cells from this culture were harvested via centrifugation, resuspended in 2 mL LB medium containing amp and cam, grown overnight at 37°C, and harvested for plasmid purification.

The collection of 24 variants used to relate relative enrichment (r.e.) to catalytic activity (see **Fig. S4D**) uses previously published data from site-directed mutagenesis studies of wildtype *E. coli* CM[46,48,63,64]. For this study, previously prepared glycerol stocks containing this library of "standard curve" variants, as well as the library of natural sequences were obtained from members of the Ranganathan laboratory.

C. Sequencing and Enrichment Analysis

Input and selected sample plasmids were prepared for multiplexed Illumina sequencing using two rounds of overlap-extension PCR with Q5 HiFi DNA Polymerase (New England Biolabs). The first round of PCR served to amplify CM DNA and append variable-length random sequences (6-9 N) to avoid base-calling failures during initial cluster identification. The second round of PCR added the adapter sequences and corresponding i5 or i7 Illumina indices. PCR amplification was performed using 5 cycles in the first round and 15 cycles in the second round with high template concentrations to minimize amplification-induced bias. The resulting PCR products were purified using the QIAquick Gel Extraction Kit and sequenced on an Illumina MiSeq instrument using a v3 paired end 600 cycle kit. The resulting paired-end reads were merged using FLASH, and sequences corresponding to the CM coding region were extracted and translated *in silico*. Only exact matches to sequences in each library were retained. Sequences with less than ten "input" reads were not considered, and sequences with zero "selected" reads were assigned a pseudo count of one to avoid division by zero. After these filtering steps, the frequency of each sequence was calculated and used to determine relative enrichment according to the following equation:

$$r.e. = log_{10}\left(\frac{f_{sel}^{x}}{f_{inp}^{x}}\right) - log_{10}\left(\frac{f_{sel}^{WT}}{f_{inp}^{WT}}\right) \qquad (19)$$

Here, $f_y^x$ is the frequency of observing sequence x in sample y. Label 'inp' refers to the input samples, 'WT' refers to the *E. coli* wildtype and 'sel' refers to the selected samples (post-24-hour growth in selective media). The error in relative enrichment was computed by first assuming that the counts are Poisson distributed, and then combining the error from each term (input vs. selected) as an independent source of error.

The normalization of the relative enrichment (norm r.e.) was found by first shifting the entire relative enrichment array such that the null sequence is placed at zero and then dividing by the difference between the relative enrichments of null and wildtype *E. coli.* Doing so sets the norm r.e. of the null sequence and the wildtype *E. coli* to 0 and 1 respectively. The error in norm r.e was found by dividing the error in r.e. by the between the r.e. of null and wildtype *E. coli.*

## Supplementary Tables

These tables can also be found in the GitHub repository for this paper listed below.

1. **Table S1: Data related to AlphaFold3, Rosetta and RMSD calculations for Natural and LRBM-constructed CM sequences**
2. **Table S2: Data related to the experimental assay for Natural and LRBM-constructed CM sequences**
3. **Table S3: Phylogenetic annotations for 1130 Natural CM sequences**

## Data and Code Availability

All data and code generated in our study is available in the following GitHub repository under the Apache 2.0 license. https://github.com/aramanlab/Pandey_et_al_2026.git

## Acknowledgements

We thank members of the David Pincus, Rama Ranganathan, Peter Littlewood, and Madhav Mani for helpful discussions. The work is supported by the National Institute of Theory and Mathematics in Biology (NITMB) and the National Institutes of Health (NIH Grant: R35GM146702).

## Author Information

C.P. performed all experiments related to evaluating synthetic and natural chorismate mutase sequences. V.R. and N.R. performed analysis related to language. B.P. conducted all theoretical advances, wrote all code, and performed all analysis. A.S.R. conceived of the approach and supervised all aspects of data collection, analysis, and experimental design. B.P. and A.S.R. wrote the manuscript.

## Ethics Declarations

The authors have nothing to declare.

## Materials and Correspondence

Author to whom correspondence and materials request should be addressed is A.S.R.